\documentclass[fleqn,usenatbib,twocolumn]{mnras}

\usepackage[T1]{fontenc}
\usepackage{ae,aecompl}
\usepackage{lineno}

\usepackage[caption=false]{subfig}
\usepackage{tikz}

\usepackage{graphicx}	
\usepackage{amsmath}	
\usepackage{amssymb}	
\usepackage{booktabs}
\usepackage{multirow}

\usepackage{newtxtext,newtxmath}

\usepackage{bm}
\usepackage{caption}

\newcommand{\Arg}[1]{{\left( #1 \right)}}

\newcommand{\Args}[1]{{\left[ #1 \right]}}
\newcommand{\Argf}[1]{{\left\{ #1 \right\}}}
\newcommand{\Avg}[1]{{\left\langle #1 \right\rangle}}
\newcommand{\Abs}[1]{{\left| #1 \right|}}

\let\vec\mathbf

\DeclareMathOperator{\cov}{cov}
\DeclareMathOperator{\tr}{tr}
\DeclareMathOperator{\Var}{Var}

\newcommand{\desimtwo}{DESI-M2}
\newcommand{\ihMpc}{h^{-1}{\rm Mpc}}

\newcommand{\EZmocksraw}{Firstgen EZ mocks}
\newcommand{\EZmockraw}{Firstgen EZ mock}
\newcommand{\EZmocks}{{\tt \EZmocksraw{}}}
\newcommand{\EZmock}{{\tt \EZmockraw{}}}
\newcommand{\themocks}{\desimtwo{} \EZmocks{}}
\newcommand{\themock}{\desimtwo{} \EZmock{}}

\newcommand{\rascalc}{{\sc RascalC}}

\numberwithin{equation}{section}

\title[2PCF covariance matrices for early DESI data]{Validation of semi-analytical, semi-empirical covariance matrices for two-point correlation function for early DESI data}

\author[M. Rashkovetskyi et al]{\parbox{\textwidth}{
Michael Rashkovetskyi$^{1}$\thanks{E-mail: \href{mailto:mrashkovetskyi@cfa.harvard.edu}{mrashkovetskyi@cfa.harvard.edu}},
Daniel J. Eisenstein$^{1}$,
Jessica Nicole Aguilar$^{2}$,
David Brooks$^{3}$,
Todd Claybaugh$^{2}$,
Shaun Cole$^{4}$,
Kyle Dawson$^{5}$,
Axel de la Macorra$^{6}$,
Peter Doel$^{3}$,
Kevin Fanning$^{7}$,
Andreu Font-Ribera$^{8}$,
Jaime E. Forero-Romero$^{9}$,
Satya Gontcho A Gontcho$^{2}$,
ChangHoon Hahn$^{10}$,
Klaus Honscheid$^{7}$,
Robert Kehoe$^{11}$,
Theodore Kisner$^{2}$,
Martin Landriau$^{2}$,
Michael Levi$^{2}$,
Marc Manera$^{8}$,
Ramon Miquel$^{12,8}$,
Jeongin Moon$^{13,14}$,
Seshadri Nadathur$^{15}$,
Jundan Nie$^{16}$,
Claire Poppett$^{2,17}$,
Ashley J. Ross$^{7}$,
Graziano Rossi$^{13}$,
Eusebio Sanchez$^{18}$,
Christoph Saulder$^{19}$,
Michael Schubnell$^{20}$,
Hee-Jong Seo$^{21}$,
Gregory Tarle$^{20}$,
David Valcin$^{21}$,
Benjamin Alan Weaver$^{22}$,
Cheng Zhao$^{23,24}$,
Zhimin Zhou$^{16}$,
and Hu Zou$^{16}$
} \vspace*{6pt} \\ 
Affiliations are listed at the end of the paper
}

\date{Accepted XXX. Received YYY; in original form ZZZ}

\pubyear{2023}

\begin{document}
\label{firstpage}
\pagerange{\pageref{firstpage}--\pageref{lastpage}}
\maketitle

\begin{abstract}
We present an extended validation of semi-analytical, semi-empirical covariance matrices for the two-point correlation function (2PCF) on simulated catalogs representative of Luminous Red Galaxies (LRG) data collected during the initial two months of operations of the Stage-IV ground-based Dark Energy Spectroscopic Instrument (DESI).
We run the pipeline on multiple effective Zel'dovich (EZ) mock galaxy catalogs with the corresponding cuts applied and compare the results with the mock sample covariance to assess the accuracy and its fluctuations.
We propose an extension of the previously developed formalism for catalogs processed with standard reconstruction algorithms.
We consider methods for comparing covariance matrices in detail, highlighting their interpretation and statistical properties caused by sample variance, in particular, nontrivial expectation values of certain metrics even when the external covariance estimate is perfect.
With improved mocks and validation techniques, we confirm a good agreement between our predictions and sample covariance.
This allows one to generate covariance matrices for comparable datasets without the need to create numerous mock galaxy catalogs with matching clustering, only requiring 2PCF measurements from the data itself.
The code used in this paper is publicly available at \url{https://github.com/oliverphilcox/RascalC}.
\end{abstract}




\section{Introduction}
\label{sec:intro}

Measurements of the large-scale structure of the Universe are one of the pillars of modern cosmology.
The two-point correlation function (2PCF) of galaxies is a particularly important statistical quantity for the large-scale structure, describing the excess probability of finding a galaxy at a given separation from another galaxy, compared to a random distribution.
Its measurements have a notable feature at the scale of baryon acoustic oscillations (BAO, first detected by \cite{BAO-discovery}).
As the corresponding comoving scale is a standard ruler with length set by sound horizon during recombination (more precisely, drag epoch), these measurements particularly constrain the expansion history of the Universe at redshifts between then and now ($z\sim 1$), providing a valuable test for cosmological models.
A more detailed overview of the methodology is provided in \cite{BAO-review}.

We have an exciting opportunity to analyze the data from the Dark Energy Spectroscopic Instrument (DESI, \cite{DESI1,DESI-instrument-overview,DESI-SV,DESI-EDR}), a highly promising 5-year Stage-IV BAO experiment for large-scale spectroscopic surveys.
For example, the sample of Luminous Red Galaxies (LRG, \cite{DESI-target-LRG}) observed during the first two months of DESI main survey operations (\desimtwo{}) yields a BAO scale measurement with 1.7\% precision \citep{DESI-DA02}, which is already comparable to the aggregate precision of 0.77\% of preceding leading surveys, BOSS and eBOSS \citep{SDSS4-eBOSS-Alam21}.

For rigorous interpretation of data, the likelihood is crucial.
Fortunately, the distribution of measured clustering statistics is well described by a multivariate Gaussian, which is fully described by mean and covariance matrix.
Unfortunately, the latter poses a serious challenge.
A fully analytical model for the covariance matrix is desirable because it can provide a fast and stable result.
However, it is very hard to construct in the case of galaxy clustering.
One of the reasons is that galaxies are high matter overdensities, evolving in a crucially nonlinear regime.
Another is the complicated effects of survey geometry and non-uniform selection.
There have been recent promising developments on covariance matrices for power spectra using perturbation theory \citep{CovaPT}.
However, power spectra results are difficult to apply to the correlation functions, because the Fourier transform is not local.

The standard approach is using a sample covariance estimated from mock (simulated) galaxy catalogs.
This solution is far from ideal.
Such catalogs need to capture key aspects of clustering and be representative of the data, or assumed theoretical model.
Detailed simulations are computationally expensive, while for precise covariance estimate a large number of samples is required, increasingly higher as more quantities are measured.
This forces a hard compromise between quality and quantity, keeping the total computation time very long.
The mock-based covariance production is therefore not characterized by flexibility, as the generation and processing of numerous simulations for an updated dataset or alternative model require a huge effort.

Another group of methods is called internal for using only the data itself.
The lack of dependence on mocks and model assumptions makes them attractive.
These are represented by re-sampling techniques like jackknife and bootstrap, which involve splitting the data into parts.
However, it may be hard to ensure that the parts are principally equivalent (so the differences all stem from random fluctuations) and separate the independent contributions to different estimates used for covariance.
\cite{MP21} attempt to mitigate the latter issue for jackknives by modifying the pair weighting for jackknife covariance estimates.
However, \cite{fitted-jk} find that the results are still prone to a density-dependent bias.

Each of the three approaches has its flaws and combining advantages from different ones seems very promising.
Thus we choose to focus on a method that started nearly analytical and employed elements of an internal approach.
\cite{rascal} demonstrated that the covariance matrix for 2PCF in arbitrary survey geometry can be expressed through integrals or sums in configuration space (this result is analogous to \cite{variance-cf-estimates}), which can be computed efficiently with importance sampling techniques.
There are terms containing higher-point correlations (3-point and connected 4-point functions), which are highly challenging to model or measure with precision.
Thus instead a procedure of leaving only Gaussian correlation and rescaling of shot noise amplitude was proposed and found to achieve a good agreement with mock-based covariances.
This rescaling parameter can be calibrated on a small suite of mocks, or on a jackknife covariance intrinsic to the data analyzed \citep{rascal-jackknife}.
We find the combination of analytical methods with jackknife more promising than with mocks, because, after a validation, such a procedure does not require the construction of any new mocks to match updated data or an alternative model with different assumptions, and does not require any more than jackknife covariance computation, at the same time offering higher smoothness, stability, and invertibility.
\cite{rascalC} introduced the \rascalc{}\footnote{\url{https://github.com/oliverphilcox/RascalC} (Oliver Philcox, Daniel Eisenstein, Ross O'Connell, Alexander Wiegand, Misha Rashkovetskyi, Yuting Wang, Ryuichiro Hada, Uendert Andrade)} code with a new algorithm, making it easier to account for survey geometry via a catalog of random points, boosting the efficiency several times and extended the formalism to covariances for multiple tracers (galaxy types).
\cite{rascalC-legendre-3} developed the estimators for Legendre-binned 2PCF and isotropic three-point correlation function.

We have contributed \rascalc{} covariances for the BAO analysis of \desimtwo{} data \citep{DESI-DA02}.
This work accompanies it, focusing on the validation of the approach in realistic circumstances.
We limit ourselves to analogs of DESI LRG sample \citep{DESI-target-LRG} due to the availability of a large suite of mocks with corresponding cuts, providing a good sample covariance matrix for reference.
Similarly to \cite{rascalC}, we process a single mock catalog in essentially the same manner as data and compare the resulting covariance with the sample covariance of clustering measurements in all available mocks, which gives a fair proxy of the pipeline performance on data and is also robust to the mismatch between data and mock clustering.
We repeat the procedure multiple times taking a different catalog each time to assess the accuracy of the method, its stability, and fluctuations.
In addition, we pay extra attention to the formation of a covariance matrix comparison toolkit.
We focus on the meaning of the numbers used and derive reference values for the ideal case when the semi-analytical prediction matches the true underlying covariance.
Due to sample variance, these expectation values can be nontrivial, and understanding the noise in the comparison measures is crucially important as well.
We choose a smaller number of observables for lower noise and clearer interpretation, and further project the covariances into the lower-dimensional and more physically meaningful space of model parameters.

We also note the prospects of standard reconstruction techniques that aim to reverse the large-scale displacements during the times after the drag epoch.
Such subsequent evolution leads to broadening and contamination of the BAO peak, thus undoing it sharpens the feature \citep{BAO-recon-improvement}.
The \rascalc{} formalism is applicable to reconstructed 2PCF covariance as well with minor adjustments.

This paper is organized in the following manner.
We review previous 2PCF estimators and covariances, discuss a modification of random counts computation and a formal extension to reconstructed data in Sec.~\ref{sec:cov-estimation}.
In Sec.~\ref{sec:cov-comparison} we discuss the problem of covariance matrix comparison and present our selection of methods,
before applying them to \rascalc{} validation with DESI LRG mocks in Sec.~\ref{sec:validation}.
We conclude in Sec.~\ref{sec:conclusion} by reviewing current accomplishments and future prospects.
Appendix~\ref{sec:previous-work-ext} provides more complete details on the covariance matrix estimators.
Appendix~\ref{sec:cov-comparison-properties} provides an overview and derivations of useful properties of covariance matrix comparison metrics.

\section{Methods of covariance matrix estimation}
\label{sec:cov-estimation}

We start by recapitulating the 2PCF estimators and covariance matrix formalism from \cite{rascal}, \cite{rascal-jackknife}, and \cite{rascalC}, with a revised notation similar to \cite{rascalC-legendre-3}.
In the following two subsections, we discuss a slight modification for optimized disjoint random count computation and an extension for reconstructed data.

\subsection{Overview of previous work}
\label{subsec:previous-work}

In a galaxy survey, we may define the 2PCF of tracers $X$ and $Y$ through the ratio of pair counts:
\begin{eqnarray} \label{eq:Landy-Szalay}
\hat\xi^{XY}(r,\mu) = \frac{N^XN^Y(r,\mu)}{R^XR^Y(r,\mu)}
\end{eqnarray}
\citep{Landy-Szalay}, where $\mu=\cos\theta$ is used instead of angle $\theta$ ($\mu$ can be restricted to $0\le\mu\le 1$ by symmetry) and $N^X=D^X-R^X$.
$D^X$ and $R^X$ are (weighted) galaxies and random particles (tracing the expected mean density) of kind $X$ respectively.
In radial bin $a$ and angular bin $c$, this estimate transforms to
\begin{eqnarray} \label{eq:Landy-Szalay-binned}
\Arg{\hat\xi^{XY}}_a^c = \frac{\Arg{N^XN^Y}_a^c}{\Arg{R^XR^Y}_a^c}
\end{eqnarray}
with
\begin{eqnarray} \label{eq:NN-RR-discrete}
\Arg{N^XN^Y}_a^c &=& \sum_{i\neq j}n^X_in^Y_jw^X_iw^Y_j\Theta^a(r_{ij})\Theta^c(\mu_{ij})\delta^X_i\delta^Y_j \\ \nonumber
\Arg{R^XR^Y}_a^c &=& \sum_{i\neq j}n^X_in^Y_jw^X_iw^Y_j\Theta^a(r_{ij})\Theta^c(\mu_{ij}),
\end{eqnarray}
where we have assigned a cubic grid to the survey such that each cell contains no more than one galaxy, $n^X_i$ is the expected mean number density of tracer $X$ in the cell $i$, $w_i$ is the expected mean weight, $\delta_i$ is the fractional galaxy overdensity, $\mu_{ij}$ relates to the angle between the line of sight and the separation vector $\vec r_{ij} = \vec r_i - \vec r_j$ ($r_{ij}$ being its absolute value), and $\Theta$ are binning functions (unity if the argument fits into the bin and zero otherwise).

Given the binned 2PCF estimator (Eq.~\eqref{eq:Landy-Szalay-binned}), the covariance matrix can be computed by definition:
\begin{eqnarray} \label{eq:2x2cov-def}
\cov \Args{\Arg{\hat\xi^{XY}}_a^c, \Arg{\hat\xi^{ZW}}_b^d} & = & \Avg{\Arg{\hat\xi^{XY}}_a^c \Arg{\hat\xi^{ZW}}_b^d} \\ \nonumber
& & - \Avg{\Arg{\hat\xi^{XY}}_a^c} \Avg{\Arg{\hat\xi^{ZW}}_b^d},
\end{eqnarray}
where $\Avg{}$ means ensemble average over realizations of overdensity $\delta$.
According to Eq.~\eqref{eq:NN-RR-discrete}, this affects the $NN$ counts but not $RR$.
Thus the right-hand side of Eq.~\eqref{eq:2x2cov-def} has the product of $RR$ counts as a constant denominator and a sum over sets of 4 cells (from the product of $NN$ counts) containing an ensemble average of the product of 4 $\delta$ values, where some of the cells can coincide with each other.
These ensemble averages are by definition 4-point correlation functions, but same-cell overdensities handled naively give zero separation and divergent values.
To overcome this issue, \cite{rascal} further expanded Eq.~\eqref{eq:2x2cov-def} into sums over distinct 2-, 3- and 4-cell configurations.
Additionally, squares of overdensity in one cell (products of overdensities in coinciding cells) were replaced by a shot-noise approximation:
\begin{eqnarray} \label{eq:shot-noise-approximation0}
\Arg{\delta^X_i}^2 \approx \frac1{n^X_i} \Arg{1+\delta^X_i}.
\end{eqnarray}
As a result, only products of 4, 3, or 2 overdensities in distinct cells remained.
After ensemble averaging, these give 4-, 3- and 2-point correlation functions at nonzero separations, respectively.
Lastly, the disconnected (Gaussian) part of the 4-point function can be separated from the connected (non-Gaussian) one according to Isserlis' (Wick's) theorem \citep{isserlis}.
The full resulting expressions are provided in Appendix~\ref{sec:previous-work-ext}, Eq.~\eqref{eq:Cov2x2_234_Point_Defs}.

Due to high noise in higher-point correlation function measurements and difficulties in their theoretical modeling, an alternative approach of mimicking non-Gaussianity has been established.
All the higher-order correlation functions are set to zero (save for disconnected 4-point, which reduces to products of 2-point functions), but the shot-noise approximation is modified with a factor $\alpha_{\rm SN}^X>1$ which can have a separate value for each of different samples of galaxies:
\begin{eqnarray} \label{eq:shot-noise-approximation}
\Arg{\delta^X_i}^2 \approx \frac{\alpha_{\rm SN}^X}{n^X_i} \Arg{1+\delta^X_i}.
\end{eqnarray}
This increases the correlations on the smallest scales, which is similar to where non-Gaussian effects are the strongest.

The sums can be transformed to continuous form by changing sums to integrals over positions and replacing cell quantities with continuous functions in 3-dimensional space.
However, it is convenient to leave them discrete, which allows us to estimate them using importance sampling directly from the random catalog \citep{rascalC}, without the need to write functional forms for survey number density, weights, and so on.
Two-point function values for pairs of points are interpolated from a grid/table to sampled pair separation $r_{ij},\mu_{ij}$ (the latter can be computed with respect to the midpoint radius-vector of the pair, or a fixed axis) with the bicubic method, which is in practice based on radially and angularly binned 2PCF estimates.
A special iterative correlation function rescaling procedure is used to modify the values of the correlation function on the interpolation grid such that the bin-averaged values of the interpolation result resemble the binned 2PCF estimates more closely.


Originally \cite{rascal} fit the shot-noise rescaling to a sample covariance obtained from a smaller set of mocks, with the idea that lower precision was sufficient for obtaining just one parameter as opposed to estimating the whole matrix.
\cite{rascal-jackknife} proposed that a jackknife covariance from the data itself can be used instead, eliminating the dependence on mocks completely.
Noting that jackknife has issues in the cosmological application, they developed a separate estimator for this covariance taking into account correlations between different estimates.

We follow a modified formalism from \cite{rascalC} called {\it unrestricted jackknife}.
According to it, the jackknife correlation function estimate $\xi_A$ is the cross-correlation function between jackknife region number $A$ and the whole survey.
In other words, an additional weight of particle $i$ (denoted by $q_i^A$) is equal to one if it belongs to the region $A$ and zero if not.
A pair of particles is additionally weighted by the mean of their weights: one if both belong to the region $A$, one half if only one does and zero if both are out of it.
Conveniently, the sum of these weights for any selected pair over all jackknife regions is unity.
As a consequence, the mean of jackknife 2PCF estimates (weighted by $RR$ counts) is equal to the full 2PCF estimate.
Then an estimator for jackknife covariance is worked out separately (Eq.~\eqref{eq:Cov2x2estimator-jack}), with a similar shot-noise rescaling procedure.

In principle, shot-noise rescaling (Eq.~\eqref{eq:shot-noise-approximation}) can be different for each tracer, and it can be obtained by fitting the prediction for the jackknife covariance of its auto-correlation function to the data-based jackknife estimate.
The resulting $\alpha^X_{\rm SN}$ value(s) is used together with the full covariance estimator (Eq.~\eqref{eq:Cov2x2estimator}) for the final result.

Let us reiterate the key approximation: non-Gaussianity can be mimicked by rescaling shot noise while dropping the terms with higher-order correlation functions.
It works because the primary effect of non-Gaussian contributions is an additional correlation at small distances, typically smaller than the bin width of 2PCF used in actual fits and requiring a covariance.
Enhancing the shot noise results in increased correlation on infinitely small scales.
This should remain a good estimate as long as the correlation functions' contribution to covariances on scales of interest is dominated by their squeezed limits.
Whether this is the case is not clear generally, but \cite{rascal} reported good agreement with large sets of mock catalogs achieved with this method, and \cite{SDSS-rascal} demonstrated the applicability of the approach to BAO analysis.


It is important to specify the means of fitting covariance matrices.
Following \cite{rascalC}, we choose to optimize the Kullback-Leibler (KL) divergence between the \rascalc{} jackknife precision ${\bf \tilde \Psi}_J \Arg{\alpha_{\rm SN}}$ and the data-based jackknife covariance estimate (given by Eq.~\eqref{eq:cov-jackknife-def}), $D_{\rm KL} \Args{{\bf \tilde \Psi}_J \Arg{\alpha_{\rm SN}}, {\bf C}_J}$, where
\begin{eqnarray} \label{eq:D_KL-gaussian0}
D_{\rm KL} \Arg{{\bf \Psi}_1, {\bf C}_2} = \frac12 \Args{\tr \Arg{{\bf \Psi}_1 {\bf C}_2} - N_{\rm bins} - \ln \det \Arg{{\bf \Psi}_1 {\bf C}_2}}.
\end{eqnarray}
In this equation, $N_{\rm bins}$ is the dimension of covariance matrices (number of correlation function bins).
Inversion of the \rascalc{} covariance has a certain bias.
Since it is not expected to obey Wishart statistics like the (mock) sample covariance, the Hartlap factor \citep{hartlap-factor} is not relevant.
A special second-order bias correction has been derived in \cite{rascal-jackknife}:
\begin{eqnarray} \label{eq:bias-correction}
{\bf \tilde \Psi} &=& ({\bf \mathbb I} - {\bf \tilde D}) {\bf \tilde C}^{-1} \\ \nonumber
{\bf \tilde D} &=& \frac{N_\mathrm{subsamples}-1}{N_\mathrm{subsamples}} \Args{-{\bf \mathbb{I}}+\frac{1}{N_\mathrm{subsamples}}\sum_{i=1}^{N_\mathrm{subsamples}} {\bf \tilde C}_{[i]}^{-1} {\bf \tilde C}_i},
\end{eqnarray}
which uses the partial covariance estimates ${\bf \tilde C}_i$ from $N_\mathrm{subsamples}$ distinct sets of configurations resulting from importance sampling in the estimation of sums (Eq.~\eqref{eq:Cov2x2_234_Point_Defs} or \eqref{eq:Cov2x2_234_Point_Defs-jack}) and mean of all the partial estimates but the $i$'th ${\bf \tilde C}_{[i]}$.
Thus the correction is applicable for both jackknife and full covariance.

\subsection{Split random-random computation}
\label{sec:split-RR}

\citet{split-randoms} showed that splitting the random catalog into a number of sub-catalogs of the same size as the data catalog when calculating random–random pairs and excluding pairs across different sub-catalogs provides the optimal error at a fixed computational cost.
The splitting can be used in \rascalc{}.
It gives little to no speed-up and impact on results because the importance sampling is too far from complete.
However, it can be useful for multi-node parallelization.
This approach has been used for the data-based \rascalc{} computation in \cite{DESI-DA02}.

A robust implementation of split random-random pair calculations in \rascalc{} would require considering only quadruples of random points where members of each pair are from the same sub-catalog, but the pairs can be from different catalogs.
However, this has been found to have little to no impact on the results, probably due to the fact that importance sampling covers only a small fraction of all possible configurations.
At the same time, such implementation makes the code less efficient and makes it impossible to split the computation of different catalogs between nodes.

\subsection{Reconstructed two-point function covariance}
\label{subsec:post-recon}

After standard reconstruction, a common approach is to replace $N=(D-R)$ by $N=(D-S)$ ($S$ being the random point with position shifted in the same manner as data) in the Landy-Szalay estimator (Eq.~\eqref{eq:Landy-Szalay}), leaving $RR$ in the denominator, so that it is
\begin{eqnarray} \label{eq:Landy-Szalay-recon}
\hat\xi^{XY} = \frac{D^XD^Y - D^XS^Y - S^XD^Y + S^XS^Y}{R^XR^Y},
\end{eqnarray}
instead of
\begin{eqnarray} \label{eq:Landy-Szalay-alt}
\hat\xi^{XY} = \frac{D^XD^Y - D^XR^Y - R^XD^Y + R^XR^Y}{R^XR^Y}.
\end{eqnarray}
This means shifted randoms are to be used in sums or integrals representing $NN$.
These eventually form the sums (or integrals) for $C$ terms.
Thus, strictly speaking, the procedure for reconstructed 2PCF should be:
\begin{itemize}
\item use shifted randoms for sampling, corresponding to the numerator of \eqref{eq:Landy-Szalay-recon};
\item provide a differently normalized 2PCF as input, namely
\begin{eqnarray} \label{eq:input-2PCF}
\hat\xi^{XY}_{\rm in} = \frac{D^XD^Y - D^XS^Y - S^XD^Y + S^XS^Y}{S^XS^Y},
\end{eqnarray}
since non-shifted randoms do not appear in the sampling procedure;
\item use non-shifted random counts for denominator in Eq.~\eqref{eq:Cov2x2_234_Point_Defs}, or correction function (Eq.~\eqref{eq:RR_approximation_generalized}) in Legendre case, corresponding to the denominator of \eqref{eq:Landy-Szalay-recon}.
\end{itemize}

Shifted randoms are individual for each mock catalog.
Therefore they can not be defined clearly for mock-averaged computations.
In those cases, we continue to use the non-shifted randoms everywhere for consistency.

\section{Methods of comparison of covariance matrices}
\label{sec:cov-comparison}

Since a covariance matrix is a high-dimensional object, it can be hard to explore and interpret.
Moreover, we run the pipeline multiple times independently and aim to study all the covariance matrix products to assess their stability and fluctuations.
Thus compact and numerical comparison measures are instructive.

\subsection{Interpretable measures of similarity for covariance matrices}
\label{subsec:comparison-measures}

The first characteristic we consider is the Kullback-Leibler (KL) divergence, a measure of distance between distributions used to fit covariances in \rascalc{} (Section~\ref{subsec:previous-work}, Eq.~\eqref{eq:D_KL-gaussian0}).
It is generally defined as an expectation value of the logarithm of the ratio of the two probability distribution functions according to the first distribution:
\begin{eqnarray} \label{eq:D_KL-definition}
D_{\rm KL} (P_1 || P_2) = \int \ln \Arg{\frac{P_1(x)}{P_2(x)}} P_1(x) dx
\end{eqnarray}
By this expression, KL divergence can be seen as an average difference in log-likelihood.
For two Gaussian distributions with covariance matrices ${\bf C}_i$ and precision matrices ${\bf \Psi}_i={\bf C}_i^{-1}$ describing $N_{\rm bins}$ observables (correlation function bins in our setup), it can be found as
\begin{eqnarray} \label{eq:D_KL-gaussian}
D_{\rm KL} \Arg{{\bf \Psi}_1, {\bf C}_2} = \frac12 \Args{\tr \Arg{{\bf \Psi}_1 {\bf C}_2} - N_{\rm bins} - \ln \det \Arg{{\bf \Psi}_1 {\bf C}_2}}.
\end{eqnarray}
This is the expression we will use.
\cite{rascal} show that the KL divergence is related to the log-likelihood if the covariance matrix is estimated from a sample with multivariate normal distribution, which is a very good approximation for correlation function bins, and additionally assuming that the precision matrix is the inverse of the true covariance matrix characterizing the multivariate normal distribution of measured quantities.
This is very appropriate for testing the hypothesis that the \rascalc{} precision matrix is a precise unbiased estimate.

The next metric assesses how close the first precision matrix is to the inverse of the second covariance matrix, and at the same time a ``directional'' root-mean-square relative difference in $\chi^2$ given by the two covariance matrices (explained in more detail in Appendix~\ref{subsec:inv-test}):
\begin{eqnarray} \label{eq:R_inv-definition}
R_{\rm inv} \Arg{{\bf \Psi}_1, {\bf C}_2} &=& \frac1{\sqrt{N_{\rm bins}}} \Abs{\Abs{{\bf C}_2^{1/2} {\bf \Psi}_1 {\bf C}_2^{1/2} - {\bf \mathbb I}}}_F \nonumber \\
&=& \sqrt{\frac{\tr \Args{\Arg{{\bf \Psi}_1 {\bf C}_2 - {\bf \mathbb I}}^2}}{N_{\rm bins}}}.
\end{eqnarray}
This measure can also be seen as the average relative difference in the errorbars.
Moreover, if the covariance matrix is estimated from a sample with multivariate normal distribution and the precision matrix is assumed to be true, $R_{\rm inv}^2$ is proportional to the $\chi^2$ computed using covariance of independent covariance matrix elements (Appendix~\ref{subsec:inv-test}).
Thus, in this case, it can serve as an approximation of log-likelihood for optimization.

The last metric is akin to the mean reduced $\chi^2$ of samples corresponding to one covariance matrix with respect to the other precision matrix:
\begin{eqnarray} \label{eq:chi2_red-definition}
\chi^2_{\rm red} \Arg{{\bf \Psi}_1, {\bf C}_2} = \frac1{N_{\rm bins}} \tr \Arg{{\bf \Psi}_1 {\bf C}_2}.
\end{eqnarray}
It can be seen as the mean ratio of $\chi^2$ given by the two covariance/precision matrices.

All three metrics are not symmetric, meaning that values for $\Arg{{\bf \Psi}_1, {\bf C}_2}$ and $\Arg{{\bf \Psi}_2, {\bf C}_1}$ may be different, so in principle, it might be informative to consider differences both ways.
On the other hand, the sample covariance is less robust than the \rascalc{} result and its inversion can be less stable.
Moreover, computing each metric twice makes the results more numerous and less clear.
Finally, the KL divergence (Eq.~\ref{eq:D_KL-gaussian}) is expected to lose its log-likelihood sense if computed between the sample precision and model covariance, since the latter one does not necessarily follow Wishart distribution.
Therefore we decided to limit ourselves to \rascalc{} precision matrices and sample covariance matrices.

For a better understanding of the metrics, let us consider the eigenvalues of ${\bf \Psi}_1 {\bf C}_2$ (alternatively, one can use ${\bf \Psi}_1^{1/2} {\bf C}_2 {\bf \Psi}_1^{1/2}$ which is symmetric) and denote them as $\lambda_a$.
We would like ${\bf \Psi}_1 \rightarrow {\bf C}_2^{-1}$ thus all $\lambda_a \rightarrow 1$.
The metrics then can be expressed as
\begin{eqnarray} \label{eq:D_KL-explanation}
D_{\rm KL} \Arg{{\bf \Psi}_1, {\bf C}_2} = \frac12 \sum_{a=1}^{N_{\rm bins}} \left[ \lambda_a - 1 - \ln \lambda_a \right] \approx \frac14 \sum_{a=1}^{N_{\rm bins}} (\lambda_a - 1)^2,
\end{eqnarray}
\begin{eqnarray} \label{eq:R_inv-explanation}
R_{\rm inv} \Arg{{\bf \Psi}_1, {\bf C}_2} = \sqrt{\frac1{N_{\rm bins}} \sum_{a=1}^{N_{\rm bins}} (\lambda_a - 1)^2},
\end{eqnarray}
\begin{eqnarray} \label{eq:chi2_red-explanation}
\chi^2_{\rm red} \Arg{{\bf \Psi}_1, {\bf C}_2} = \frac1{N_{\rm bins}} \sum_{a=1}^{N_{\rm bins}} \lambda_a.
\end{eqnarray}
Thus $D_{\rm KL}$ and $R_{\rm inv}$ accumulate any deviation of $\lambda_a$ from 1, although they cannot indicate the direction of such differences.
Note that the quadratic expression for $D_{\rm KL}$ is approximate so it is not generally degenerate with $R_{\rm inv}$, although as the covariance matrices approach each other these two measures become more redundant:
\begin{eqnarray} \label{eq:D_KL-R_inv-redundancy}
D_{\rm KL} \Arg{{\bf \Psi}_1, {\bf C}_2} \approx \frac{N_{\rm bins}}4 R_{\rm inv}^2 \Arg{{\bf \Psi}_1, {\bf C}_2}.
\end{eqnarray}
$\chi^2_{\rm red}$ can show which covariance matrix is ``larger'' on average, while deviations in opposite directions may cancel each other.

Next, we would like to understand what to expect from these metrics.
For this purpose, we consider the case when the precision matrix is predicted perfectly (${\bf \Psi}_0$), ideally matching the true underlying covariance (${\bf C}_0={\bf \Psi}_0^{-1}$), and focus on the noise properties of the sample covariance matrix ${\bf C}_S$ obtained via the standard unbiased estimator for the case when the true mean is not known:
\begin{eqnarray} \label{eq:sample-cov}
C_{S,ab} = \frac1{n_S-1} \sum_{i=1}^{n_S} (\xi_{a,i} - \bar \xi_a) (\xi_{b,i} - \bar \xi_b)
\end{eqnarray}
where $a,b$ denote bin numbers, $i,j$ index sample numbers, and $\bar \xi_a$ is the estimate of the mean:
\begin{eqnarray}
\bar \xi_a \equiv \frac1{n_S} \sum_{i=1}^{n_S} \xi_{a,i}.
\end{eqnarray}
Since the clustering measurements are described well by a multivariate normal distribution, their sample covariance matrix follows the Wishart statistics.
This provides a reference of how the metrics behave when the perfect precision matrix is compared to a covariance matrix estimated from $n_S$ samples with $N_{\rm bins}$ bins (or any other Gaussian observables).
Full derivations are presented in Appendix~\ref{sec:cov-comparison-properties}, here we will only provide the results for mean/expectation values and standard deviations:
\begin{eqnarray} \label{eq:D_KL-stats}
& \Avg{D_{\rm KL} \Arg{{\bf \Psi}_0, {\bf C}_S}} \approx \frac{N_{\rm bins}(N_{\rm bins}+1)}{4(n_S-1)}, \nonumber \\
& \sigma \Args{D_{\rm KL} \Arg{{\bf \Psi}_0, {\bf C}_S}} \approx \frac12 \sqrt{\frac{N_{\rm bins} [(N_{\rm bins} + 1) (n_S + 2 N_{\rm bins} + 2) + 2]}{\Arg{n_S-1}^3}};
\end{eqnarray}
\begin{eqnarray} \label{eq:R_inv-stats}
& \Avg{R_{\rm inv} \Arg{{\bf \Psi}_0, {\bf C}_S}} \approx \sqrt{\frac{N_{\rm bins}+1}{n_S-1}}, \nonumber \\
& \sigma \Args{R_{\rm inv} \Arg{{\bf \Psi}_0, {\bf C}_S}} \approx \frac1{n_S-1} \sqrt{\frac{(N_{\rm bins} + 1) (n_S + 2 N_{\rm bins} + 2) + 2}{N_{\rm bins} (N_{\rm bins}+1)}}.
\end{eqnarray}
Naively, one could expect $D_{\rm KL}$ and $R_{\rm inv}$ to become arbitrarily small as ${\bf \Psi}_1 \rightarrow {\bf \Psi}_0$.
However, in reality, they can have large expectation values, especially as the number of bins increases.

$\chi^2_{\rm red}$, however, would behave like the reduced $\chi^2$ with $N_{\rm bins} \times (n_S-1)$ degrees of freedom in this case (see Appendix~\ref{subsec:chi2}):
\begin{eqnarray} \label{eq:chi2_red-stats}
& \Avg{\chi^2_{\rm red} \Arg{{\bf \Psi}_0, {\bf C}_S}} = 1, \nonumber \\
& \sigma \Args{\chi^2_{\rm red} \Arg{{\bf \Psi}_0, {\bf C}_S}} = \sqrt{\frac2{N_{\rm bins} (n_S-1)}}.
\end{eqnarray}
It might seem like $D_{\rm KL}$ and $R_{\rm inv}$ could be unbiased by multiplying one of the matrices by a factor similar to the Hartlap factor \citep{hartlap-factor}, but a lack of bias in $\chi^2_{\rm red}$ expectation value suggests that this is not true.

It is notable that a deviation of $\chi^2_{\rm red}$ from 1 would contribute to $R_{\rm inv}$ -- see expressions \eqref{eq:chi2_red-explanation} and \eqref{eq:R_inv-explanation}; their consequence is also that
\begin{eqnarray} \label{eq:chi2_red-R_inv-limit}
\Abs{\chi^2_{\rm red} \Arg{{\bf \Psi}_1, {\bf C}_2} - 1} \le R_{\rm inv} \Arg{{\bf \Psi}_1, {\bf C}_2}.
\end{eqnarray}
However, the direction of such deviation will not be clear in $R_{\rm inv}$, and nontrivial expectation value can make it harder to interpret.
This keeps $\chi^2_{\rm red}$ useful in many cases.

\subsection{Internal convergence assessment}
\label{subsec:internal-convergence-metrics}

Internal consistency of \rascalc{} covariance matrices in one run and the convergence of the Monte-Carlo integration procedure are also important to assess quantitatively.
We propose to employ the above-mentioned methods to accomplish this and provide valuable diagnostics which do not rely on a reference (e.g. sample) covariance and thus can be used in any run, including the pure data-based one.
However, we need to note that such a test can only quantify limited sources of uncertainty or error, leaving aside the factors like adequacy of the approximations in the formalism, the precision of the input clustering, and noise in the jackknife covariance estimated from the data.

The \rascalc{} code provides multiple partial intermediate results corresponding to practically non-overlapping sets of quadruples, triples, and pairs of points.
These resulting covariance matrices can be split into two distinct sets of similar size, averaged within them, and compared using the three metrics.
In this case, however, the arguments for the $\chi^2_{\rm red}$ weaken -- we can expect $R_{\rm inv}$ to become arbitrarily low as the number of Monte-Carlo samples increases, which would limit the reduced chi-squared via Eq.~\eqref{eq:chi2_red-R_inv-limit}, and it is not as interesting to understand which of the halves gives a ``smaller'' matrix.
Then $D_{\rm KL}$ also becomes more redundant with $R_{\rm inv}$ via Eq.~\eqref{eq:D_KL-R_inv-redundancy}.
Therefore it is reasonable to only show $R_{\rm inv}$, which can be seen as an estimate of root-mean-square relative precision (considered over all directions in measurement space).

\section{Application to DESI LRG mocks}
\label{sec:validation}

In this section, we use the described methods on \desimtwo{} mocks to assess the performance and stability of the approach on the actual dataset.
We describe the setup first, then perform intrinsic validation described in Section~\ref{subsec:internal-convergence-checks}, look at the shot-noise rescaling values resulting from jackknife calibration used for the final covariance estimates, validate the \rascalc{} results by comparison with the mock sample covariance in measurement/observable and parameter space and finally focus specifically on errorbars on BAO scale.

\subsection{Mock catalogs and reconstruction method}
\label{subsec:mocks}

We use the 999 effective Zel'dovich (EZ) mocks \citep{EZmocks,EZmocks2021} with cuts corresponding to the DESI LRG sample \citep{DESI-target-LRG} (described in more detail in \cite{DESI-DA02}), which will be referred to as \themocks{}.
Sample covariance based on these does not provide a perfect reference because both the number of mock catalogs and the level of details in each simulation are limited, but the best one can have realistically since increasing one without making the other worse would require even more significant computational resources.
Comparing these is also robust to the mismatch between data and mock clustering.

The reconstruction method is also the same as in \cite{DESI-DA02}: the iterative procedure \citep{recon-fourier-space} implemented in the {\tt IterativeFFTReconstruction} 
algorithm of the {\sc pyrecon} package\footnote{\url{https://github.com/cosmodesi/pyrecon} (Arnaud de Mattia, Martin J. White, Julian E. Bautista, Pedro Rangel Caetano, Sesh Nadathur, Enrique Paillas, Grant Merz, Davide Bianchi)} with the {\tt RecIso} convention.
Three iterations are used with a Gaussian smoothing kernel of width $15 \ihMpc$. 
An approximate growth rate and the expected bias are assumed.

\subsection{Setup}
\label{subsec:validation-setup}

For this study, we have performed separate runs using 2PCF measured from single LRG \themocks{} catalogs.
This has been repeated 10 times for pre- and post-recon.
In the latter case, individual shifted random catalogs have been used for each mock following the procedure we described in Section~\ref{subsec:post-recon}.

Pre-reconstruction galaxies and randoms were assigned unity weights, for post-reconstruction FKP weights \citep{FKP} were used, given by
\begin{equation}
    w_{\rm FKP} = \frac{1}{1+n(z) C P_0}
\end{equation}
where $n(z)$ is the weighted number density (per volume),
$C$ is the mean completeness for the sample,
and $P_0$ is a fiducial power-spectrum amplitude.
For LRG, $C=0.579$ and $P_0=10^4 (\ihMpc)^3$ \citep{DESI-DA02}.
We note that the weighting schemes are not exactly the same as for real data, but since weights are included explicitly in the covariance estimators we expect \rascalc{} to work with any fixed choice applied consistently for 2PCF measurements and Monte-Carlo integration.

For the importance sampling input, 10 random catalogs were used in pre-recon computations and 20 in post-recon, like for the $RR$ ($SS$) pair counts computation for the 2PCF estimates.
These randoms have been concatenated before being provided to \rascalc{} executable.
We assign 60 jackknife regions assigned by a $K$-means subsampler based on data positions (but not weights) as in \desimtwo{} data, compute the jackknife covariance matrix and use it to calibrate the shot-noise rescaling.

We note that some validation has been performed in \cite{DESI-DA02}: \rascalc{} covariance matrix based on 2PCF averaged over all LRG \themocks{} with no shot-noise rescaling applied (and using unshifted randoms in the post-reconstruction case) has been compared to the sample covariance matrix in terms of $\chi^2$ of BAO fits, best-fit values and standard deviations of BAO isotropic scale parameter $\alpha$, yielding a good agreement.
However, there are significant limitations to this approach:
\begin{itemize}
\item effect of noise in the input clustering is significantly smaller in 2PCF averaged over $\approx 1000$ mocks than in the real data, which is close to a single mock catalog;
\item possible differences in other parameters of the BAO model or more generic aspects of correlation function have not been assessed.
\end{itemize}
Single-mock runs address the first issue since each of them is a fair proxy of the data.
The use of covariance matrix comparison metrics from Section~\ref{subsec:comparison-measures} expands on the second one.
We keep the result with mock-average 2PCF, labeled ``Average G'' (Gaussian), to assess the importance of precision of input clustering.

We also consider a shot-noise-rescaled version of the run with mock-averaged clustering, labeled ``Average NG'' (non-Gaussian).
We note that calibration of an all-mocks run on a jackknife estimate can be ambiguous or require a repeated computation of all the pair counts with jackknives, which was not done before because jackknives are not necessary for the sample covariance.
Thus we choose to fit the full covariance matrix to the mock sample covariance by minimizing the KL divergence between them (analogously to the jackknife procedure described in Section~\ref{subsec:previous-work} and Eq.~\eqref{eq:D_KL-gaussian0}).
Since only one parameter is varied in the fit, a perfect agreement is still not guaranteed.
On the other hand, this setup is clearly idealized and would be closer to the closest possible match to the mock covariance the \rascalc{} method can provide.
It comprises another useful reference to compare to the data-like performance on single mocks.

We only consider 45 radial bins, spanning 4~$\ihMpc$ each from 20 to 200~$\ihMpc$.
The \rascalc{} covariances are produced with single angular bins, which is a simplifying assumption since treating monopole more precisely in Legendre mode with shot-noise rescaling would require 2 runs per dataset, as explained in Appendix~\ref{subsec:Legendre-cov-est}.
The 2PCF measurements for the mock sample covariance use a more precise monopole estimate provided by {\sc pycorr}\footnote{\url{https://github.com/cosmodesi/pycorr} (Arnaud de Mattia, Lehman Garrison, Manodeep Sinha, Davide Bianchi, Svyatoslav Trusov, Enrique Paillas, Seshadri Nadathur, Craig Warner, James Lasker)}.
We also project them into a BAO model parameter space using the derivatives near the best fit (Fisher forecast).
The model uses only a separation range from 48 to 148~$\ihMpc$.

\subsection{Internal convergence checks}
\label{subsec:internal-convergence-checks}

We perform an intrinsic diagnostic procedure (as described in Section~\ref{subsec:internal-convergence-metrics}) to ensure that \rascalc{} integrals converged well in each run and exclude importance sampling random noise from significant error factors.
We found that pre-recon mock 6 and post-recon mock 5 showed significantly worse consistency than all the rest.
Therefore we have run them twice longer.

After that, all the \rascalc{} results have reached a high and quite uniform level of internal consistency, as presented in Table~\ref{tab:internal-convergence}.
We only show $R_{\rm inv}$, which are easier to interpret as the root-mean-square relative deviation between different partial estimates of the covariance matrix (considered over all directions in measurement space).
$\chi^2_{\rm red}$ are limited via Eq.~\eqref{eq:chi2_red-R_inv-limit}, and $D_{\rm KL}$ values are quite close to estimates from Eq.~\eqref{eq:D_KL-R_inv-redundancy}.
Due to high consistency in measurement space, we have not performed projection to parameter space here.
The splitting of Monte-Carlo subsamples has been done in a few different ways and the non-symmetric metric has been computed both ways ($\Psi_1C_2$ and $\Psi_2C_1$), but all the values were very close\footnote{In the case of disjoint running (Sec.~\ref{sec:split-RR}) there can be a meaningful difference between splittings within or between different random sub-catalogs, due to fluctuations in pair counts between these. But here all the randoms have been concatenated together.} and thus have been averaged to one number for each metric.
These low (sub-percent) internal deviations give us confidence that the Monte-Carlo integration procedure in \rascalc{} has converged well and it will not be a significant error source in further comparison.
After ascertaining this, we have not touched the covariance matrix products to be fair -- with real survey and no mocks, other validation procedures described in this paper are not available.

\begin{table}
\begin{tabular}{|c|c|c|c|}
\hline
Mock no. & $R_{\rm inv}$ pre & $R_{\rm inv}$ post \\
\hline
Average G & $1.3 \times 10^{-3}$ & $3.0 \times 10^{-3}$ \\
Average NG & $1.2 \times 10^{-3}$ & $8.1 \times 10^{-3}$ \\
\hline
1 & $3.7 \times 10^{-3}$ & $2.3 \times 10^{-3}$ \\
2 & $3.1 \times 10^{-3}$ & $2.2 \times 10^{-3}$ \\
3 & $3.8 \times 10^{-3}$ & $2.1 \times 10^{-3}$ \\
4 & $6.3 \times 10^{-3}$ & $1.9 \times 10^{-3}$ \\
5 & $4.4 \times 10^{-3}$ & $1.3 \times 10^{-3}$ \\
6 & $4.4 \times 10^{-3}$ & $1.9 \times 10^{-3}$ \\
7 & $3.7 \times 10^{-3}$ & $2.0 \times 10^{-3}$ \\
8 & $3.3 \times 10^{-3}$ & $1.7 \times 10^{-3}$ \\
9 & $3.5 \times 10^{-3}$ & $2.5 \times 10^{-3}$ \\
10 & $3.5 \times 10^{-3}$ & $2.2 \times 10^{-3}$ \\
\hline
\end{tabular}
\caption{Intrinsic pre- and post-reconstruction convergence test results in measurement space.
$R_{\rm inv}$ estimate root-mean-square relative precision, averaged over all different directions in the measurement space.
The numbers provided here are for the full covariance, but the consistency levels of the jackknife covariance prediction from \rascalc{} are similar.
They demonstrate sub-percent stability in \rascalc{} integrals and ensure that random noise in importance sampling is not a significant source of error, as these numbers are smaller than deviations observed in further comparisons (Tables~\ref{tab:comparison-measurement-pre}--\ref{tab:comparison-parameter-post}).}
\label{tab:internal-convergence}
\end{table}

\subsection{Shot-noise rescaling values}
\label{subsec:shot-noise-rescaling}

Next, we look into the shot-noise rescaling values because the final covariance estimates (with approximate non-Gaussianity) are based on them.
The shot-noise rescaling values for single mocks are obtained by fitting the separate \rascalc{} jackknife covariance prediction to the jackknife covariance estimate for each mock.
For the mock-average clustering, the full \rascalc{} covariance was fit to the mock sample covariance instead, as discussed in Section~\ref{subsec:validation-setup}.

The shot-noise rescaling values are gathered in Table~\ref{tab:shot-noise-rescalings}.
We note that all of them are greater than one (which corresponds to purely Gaussian covariance), in accordance with our expectation that the non-Gaussianity expands the errorbars.
Moreover, the mean shot-noise rescaling of the 10 single mocks is $\approx 4$ standard deviations larger than 1 both before and after reconstruction.
The values obtained from jackknife and mock covariance are consistent.
After reconstruction, the shot-noise rescaling decreases for every mock.
The pre-recon mean is larger than the post-recon one by $\approx 1.8$ standard deviations.
The scatter after reconstruction is also smaller than before.
These deviations can be caused by the random fluctuations in the input 2PCF estimates, noise in jackknife covariances, and differences in shifted randoms (for post-recon only).

The key conclusion is that we have obtained the shot-noise rescaling parameter for a data-like setup (single mock runs) with a percent-level precision.
This maps into a similar or smaller relative deviation in the rescaled covariance matrices since the 2-point term has the strongest scaling, $\propto \alpha_{\rm SN}^2$, and the 4-point term remains the same (Eq.~\eqref{eq:Cov2x2estimator1}).

\begin{table}
\begin{tabular}{ccc}
\hline
Mock no. & Pre-recon $\alpha_{\rm SN}$ & Post-recon $\alpha_{\rm SN}$ \\
\hline
Average NG & 1.096 & 1.038 \\
\hline
1 & 1.096 & 1.062 \\
2 & 1.074 & 1.043 \\
3 & 1.040 & 1.034 \\
4 & 1.077 & 1.051 \\
5 & 1.079 & 1.033 \\
6 & 1.089 & 1.030 \\
7 & 1.080 & 1.041 \\
8 & 1.102 & 1.058 \\
9 & 1.116 & 1.033 \\
10 & 1.080 & 1.041 \\
\hline
1-10 mean$\pm$std & $1.083\pm 0.020$ & $1.043\pm 0.011$ \\
\hline
\end{tabular}
\caption{Shot-noise rescaling values for the 10 mocks, on which the final covariance predictions are based, pre- and post-reconstruction.
The ``Average NG'' value is fit to mock sample covariance, and the 1-10 are fit to jackknife covariance estimates from single mocks, and the resulting numbers are consistent.
The standard deviation for single mocks is $\approx 2\%$ before and $\approx 1\%$ after reconstruction, which translate into a similar effect on relative precision of the final covariance matrices due to rescaling.}
\label{tab:shot-noise-rescalings}
\end{table}

\subsection{Measurement-space validation}

Now we proceed to comparison with the sample covariance matrices as reference, keeping in mind they are not devoid of noise so not all the comparison measures can be ideal.
We consider the higher-dimensional space of observables first, where the effects of sample variance are quite significant.
It consists of 45 bins of 2PCF, spanning 20--200~$\ihMpc$ linearly with a bin width of 4~$\ihMpc$.

Sample covariances for all original bins have been estimated using $n_S=999$ 2PCF measurements from all the \themocks{} and the standard unbiased estimator (Eq.~\eqref{eq:sample-cov}).
The procedures have been similar for pre- and post-recon.

\begin{table}
\begin{tabular}{|c|c|c|c|}
\hline
Mock no. & $D_{\rm KL} \Arg{{\bf \Psi}_R, {\bf C}_S}$ & $R_{\rm inv} \Arg{{\bf \Psi}_R, {\bf C}_S}$ & $\chi^2_{\rm red} \Arg{{\bf \Psi}_R, {\bf C}_S}$ \\
\hline
Average G & 0.793 & 0.2922 & 1.1422 \\
Average NG & 0.537 & 0.2136 & 0.9906 \\
\hline
1 & 0.721 & 0.2252 & 0.9600 \\
2 & 0.602 & 0.2234 & 0.9992 \\
3 & 0.571 & 0.2329 & 1.0477 \\
4 & 0.548 & 0.2191 & 0.9991 \\
5 & 0.755 & 0.2302 & 0.9830 \\
6 & 0.727 & 0.2315 & 0.9723 \\
7 & 0.695 & 0.2254 & 0.9675 \\
8 & 0.530 & 0.2085 & 0.9695 \\
9 & 0.610 & 0.2140 & 0.9291 \\
10 & 0.518 & 0.2106 & 0.9895 \\
\hline
1-10 & $0.628 \pm 0.089$ & $0.2221 \pm 0.0087$ & $0.982 \pm 0.031$ \\
\hline
Perfect $\Psi$ & $0.519 \pm 0.024$ & $0.2147 \pm 0.0049$ & $1.0000 \pm 0.0067$ \\
\hline
\end{tabular}
\caption{Results of general measurement-space comparison between the \rascalc{} results ($R$) and mock sample covariance ($S$) before reconstruction.
The last row provides the perfect-case reference -- expectation values and standard deviations for the three metrics, if the \rascalc{} precision matrix truly described the distribution of the mock correlation functions.
}
\label{tab:comparison-measurement-pre}
\end{table}

\begin{table}
\begin{tabular}{|c|c|c|c|}
\hline
Mock no. & $D_{\rm KL} \Arg{{\bf \Psi}_R, {\bf C}_S}$ & $R_{\rm inv} \Arg{{\bf \Psi}_R, {\bf C}_S}$ & $\chi^2_{\rm red} \Arg{{\bf \Psi}_R, {\bf C}_S}$ \\
\hline
Average G & 0.62 & 0.247 & 1.0653 \\
Average NG & 0.57 & 0.225 & 1.0028 \\
\hline
1 & 0.94 & 0.256 & 0.9463 \\
2 & 0.63 & 0.229 & 0.9725 \\
3 & 0.72 & 0.266 & 1.0031 \\
4 & 0.61 & 0.222 & 0.9657 \\
5 & 0.62 & 0.229 & 0.9924 \\
6 & 0.65 & 0.231 & 0.9817 \\
7 & 0.62 & 0.226 & 0.9706 \\
8 & 0.61 & 0.222 & 0.9631 \\
9 & 0.77 & 0.281 & 1.0214 \\
10 & 0.88 & 0.314 & 1.0162 \\
\hline
1-10 & $0.70 \pm 0.12$ & $0.247 \pm 0.031$ & $0.983 \pm 0.024$ \\
\hline
Perfect $\Psi$ & $0.519 \pm 0.024$ & $0.2147 \pm 0.0049$ & $1.0000 \pm 0.0067$ \\
\hline
\end{tabular}
\caption{Results of general measurement-space comparison between \rascalc{} results ($R$) and mock sample covariance ($S$) after reconstruction.
The last row provides the perfect-case reference -- expectation values and standard deviations for the three metrics, if the \rascalc{} precision matrix truly described the distribution of the mock correlation functions.
}
\label{tab:comparison-measurement-post}
\end{table}

The comparison measures between the \rascalc{} precision matrices (estimated via Eq.~\ref{eq:bias-correction}) and the sample covariance matrices have been computed and are presented in Table~\ref{tab:comparison-measurement-pre} for pre-reconstruction and Table~\ref{tab:comparison-measurement-post} for post-reconstruction.
First of all, there are fluctuations in the comparison measures involving the single-mock results, stemming from the input 2PCF estimates, jackknife covariances, and differences in shifted randoms (for post-recon only) -- the same causes as for scatter in $\alpha_{\rm SN}$ discussed in Section~\ref{subsec:shot-noise-rescaling}.
The individual pre-reconstruction covariances appear to agree with the mock sample covariance better than the post-reconstruction.
The covariance with mock-averaged clustering and no shot-noise rescaling, on the contrary, gives a closer agreement after reconstruction.
Before reconstruction, any individual shot-noise rescaled covariance shows better agreement than the Gaussian mock-averaged clustering run; after reconstruction, it is very often worse.
With mock-averaged clustering, the shot-noise rescaling is clearly beneficial for pre-reconstruction and less so for post-reconstruction.
This may be a hint that the shot-noise rescaling might not be doing as well after reconstruction as before.

Compared to the perfect case, \rascalc{} typically performs worse (higher $D_{\rm KL}$ and $R_{\rm inv}$, reduced chi-squared further from one), which we expect since the code (and the mocks) involve (different) approximations.
We note that on average the comparison metrics are within a couple of standard deviations of the expectation value for the true underlying covariance matrix.
On the other hand, it is the larger standard deviation in \rascalc{} results that is allowing this conclusion, and reducing the noise factors causing it (input 2PCF fluctuations, single jackknife covariance) may allow us to reach a closer agreement in the future works.

\subsection{Parameter-space validation}

In this section, we project the covariance into a lower-dimensional and more physically meaningful space of BAO model parameters.
Lower dimensionality makes the reference values for comparison metrics clearer and the results become easier to interpret.
In addition, the correlation function modes that are not physically possible or do not affect the parameter constraints are removed from consideration, which leaves only real and important ``directions'' for consideration.

We choose a commonly used BAO model\footnote{\url{https://github.com/cosmodesi/BAOfit\_xs/} (Ashley J. Ross, Juan Mena Fernandez)} \citep{SDSS3-Ross17,BOSS-DR14-Ata18} with a scalable template $\xi_0$ and three nuisance polynomial terms:
\begin{eqnarray} \label{eq:BAO-model}
    \xi_{\rm mod}(r) = B\xi_0(\alpha_{\rm BAO} r) + A_0 + A_1/r + A_2/r^2,
\end{eqnarray}
comprising $N_{\rm pars}=5$ parameters: $B, A_0, A_1, A_2, \alpha_{\rm BAO}$.

Instead of performing full fits, we use Fisher matrix formalism.
This can be seen as less precise than full fits on every mock or MCMC using a 2PCF likelihood.
On the other hand, the parameter distribution is not Gaussian when the model is not a linear function of parameters, which makes linear approximation within Fisher matrix formalism more suitable for the comparison methods we have discussed.

We estimate the parameter covariance matrix as the inverse of the Fisher matrix:
\begin{eqnarray} \label{eq:parameter-covariance}
{\bf C}_S^{\rm par} = \frac{n_S-1}{n_S-N'_{\rm bins}+N_{\rm pars}-1} \Args{{\bf M} \Arg{{\bf C'}^{\rm meas}_S}^{-1} {\bf M}^T}^{-1}
\end{eqnarray}
where the measurement-space mock sample covariance matrix ${C'}^{\rm meas}_S$ is cut to the $N'_{\rm bins}=25$ bins spanning separations from 48 to 148~$\ihMpc$ used in BAO fits, and ${\bf M}$ is the matrix of derivatives of binned 2PCF vector $\bm \xi$ with respect to parameters $\bm p$:
\begin{eqnarray}
M_{ca} \equiv \frac{\partial \xi_a}{\partial p_c}.
\end{eqnarray}
The derivatives have been taken at the best-fit parameters for the mock-averaged clustering measurements (separate before and after reconstruction).

Note that Eq.~\eqref{eq:parameter-covariance} is scaled by a correction factor according to Eq.~(B6) in \cite{density-split-clustering} to account for biases caused by both matrix inversions.
This provides an unbiased (although not noiseless) estimate of the true underlying covariance in parameter space as validated in Appendix~\ref{subsec:stats-validation}.

A similar but simpler procedure was performed with \rascalc{} products:
\begin{eqnarray} \label{eq:rascalc-parameter-prec}
{\bf \Psi}^{\rm par}_{R,cd} \approx {\bf M} \Arg{{\bf C'}^{\rm meas}_R}^{-1} {\bf M}^T,
\end{eqnarray}
where ${\bf C'}^{\rm meas}_R$ was also cut to the 25 bins spanning separations from 48 to 148~$\ihMpc$ used in BAO fits.
There is a bias correction matrix $\bf D$ for \rascalc{} (Eq.~\eqref{eq:bias-correction}), but for the results presented here absolute values of its eigenvalues are $\lesssim 10^{-3}$ thus we have decided to neglect this correction factor.

\begin{table}
\begin{tabular}{|c|c|c|c|}
\hline
Mock no. & $D_{\rm KL} \Arg{{\bf \Psi}_R, {\bf C}_S}$ & $R_{\rm inv} \Arg{{\bf \Psi}_R, {\bf C}_S}$ & $\chi^2_{\rm red} \Arg{{\bf \Psi}_R, {\bf C}_S}$ \\
\hline
Average G & 0.026 & 0.138 & 0.992 \\
Average NG & 0.027 & 0.135 & 0.925 \\
\hline
1 & 0.104 & 0.236 & 0.855 \\
2 & 0.065 & 0.193 & 0.911 \\
3 & 0.022 & 0.122 & 0.941 \\
4 & 0.013 & 0.099 & 0.962 \\
5 & 0.127 & 0.250 & 0.883 \\
6 & 0.091 & 0.232 & 0.866 \\
7 & 0.095 & 0.229 & 0.861 \\
8 & 0.025 & 0.128 & 0.918 \\
9 & 0.042 & 0.165 & 0.876 \\
10 & 0.016 & 0.107 & 0.944 \\
\hline
1-10 & $0.060 \pm 0.042$ & $0.176 \pm 0.059$ & $0.902 \pm 0.039$ \\
\hline
Perfect $\Psi$ & $0.0075 \pm 0.0028$ & $0.078 \pm 0.014$ & $1.000 \pm 0.020$ \\
\hline
\end{tabular}
\caption{Results of BAO parameter space ($B, A_0, A_1, A_2, \alpha_{\rm BAO}$) comparison between \rascalc{} results ($R$) and mock sample covariance ($S$) before reconstruction.
The last row provides the perfect-case reference -- expectation values and standard deviations for the three metrics, if the \rascalc{} precision matrix truly described the distribution of the mock correlation functions.
}
\label{tab:comparison-parameter-pre}
\end{table}

\begin{table}
\begin{tabular}{|c|c|c|c|}
\hline
Mock no. & $D_{\rm KL} \Arg{{\bf \Psi}_R, {\bf C}_S}$ & $R_{\rm inv} \Arg{{\bf \Psi}_R, {\bf C}_S}$ & $\chi^2_{\rm red} \Arg{{\bf \Psi}_R, {\bf C}_S}$ \\
\hline
Average G & 0.020 & 0.134 & 1.087 \\
Average NG & 0.014 & 0.113 & 1.049 \\
\hline
1 & 0.122 & 0.252 & 0.907 \\
2 & 0.036 & 0.171 & 1.002 \\
3 & 0.041 & 0.202 & 1.082 \\
4 & 0.015 & 0.106 & 0.981 \\
5 & 0.030 & 0.149 & 0.999 \\
6 & 0.029 & 0.144 & 0.974 \\
7 & 0.021 & 0.125 & 0.981 \\
8 & 0.014 & 0.105 & 0.977 \\
9 & 0.095 & 0.329 & 1.154 \\
10 & 0.096 & 0.327 & 1.155 \\
\hline
1-10 & $0.050 \pm 0.039$ & $0.191 \pm 0.085$ & $1.021 \pm 0.082$ \\
\hline
Perfect $\Psi$ & $0.0075 \pm 0.0028$ & $0.078 \pm 0.014$ & $1.000 \pm 0.020$ \\
\hline
\end{tabular}
\caption{Results of BAO parameter space ($B, A_0, A_1, A_2, \alpha_{\rm BAO}$) comparison between \rascalc{} results ($R$) and mock sample covariance ($S$) after reconstruction.
The last row provides the perfect-case reference -- expectation values and standard deviations for the three metrics, if the \rascalc{} precision matrix truly described the distribution of the mock correlation functions.
}
\label{tab:comparison-parameter-post}
\end{table}

The comparison measures have been computed between the projected matrices and are presented in Table~\ref{tab:comparison-parameter-pre} for pre-recon and Table~\ref{tab:comparison-parameter-post} for post-recon.
Generally, lower expectation values of $D_{\rm KL}$ and $R_{\rm inv}$ for perfect precision make these numbers for \rascalc{} easier to interpret.
There is a less apparent difference between pre- and post-recon.
A notable exception is that all $\chi^2_{\rm red}$ for rescaled (with mock-averaged and single-mock clusterings) pre-recon are significantly less than 1 (meaning \rascalc{} ``overestimates'' the covariance then).
In other cases, mimicking non-Gaussianity gives a slight improvement for mock-averaged clustering, but higher noise in 2PCF and jackknife covariance in single-mock estimates often drives the agreement with mock sample covariance worse than in the mock-averaged Gaussian estimate.

Overall, \rascalc{} single-mock results are within $<2\sigma$ (dominated by the standard deviation of the perfect reference values, except the reduced chi-squared before reconstruction, which deviates by $\approx 2.2$ std (combined).
However, the scatter in these numbers is quite significant (e.g. a few percent in root-mean-square relative error $R_{\rm inv}$), and we should try to reduce it in future work.

%

\subsection{Errorbars on BAO scale parameter}

Since the scale parameter $\alpha_{\rm BAO}$ is the important output of the current BAO analysis (Eq.~\ref{eq:BAO-model}), we have decided to extract its errorbar, marginalized over the other four parameters.
This is quite trivial after the previous subsection -- we only needed to invert the \rascalc{} parameter-space precisions
\begin{eqnarray} \label{eq:rascalc-parameter-cov}
{\bf C}^{\rm par}_R = \Arg{{\bf \Psi}^{\rm par}_R}^{-1}
\end{eqnarray}
neglecting the inversion bias, since it is expected to be even smaller than before with the smaller size of the matrices.
Then we extract the marginalized errorbars from all the parameter covariances as
\begin{eqnarray}
\sigma(\alpha_{\rm BAO}) = \sqrt{C^{\rm par}_{\alpha\alpha}}.
\end{eqnarray}

\begin{table}
\begin{tabular}{ccc}
\hline
$\sigma(\alpha_{\rm BAO})$ & Pre-recon & Post-recon \\
\hline
Sample cov & $0.01590 \pm 0.00036$ & $0.01459 \pm 0.00033$ \\
\hline
Average G & 0.01522 & 0.01360 \\
Average NG & 0.01596 & 0.01392 \\
\hline
1 & 0.01616 & 0.01424 \\
2 & 0.01598 & 0.01395 \\
3 & 0.01583 & 0.01410 \\
4 & 0.01609 & 0.01405 \\
5 & 0.01595 & 0.01385 \\
6 & 0.01613 & 0.01396 \\
7 & 0.01594 & 0.01389 \\
8 & 0.01621 & 0.01400 \\
9 & 0.01652 & 0.01402 \\
10 & 0.01618 & 0.01405 \\
\hline
1-10 mean$\pm$std & $0.01610 \pm 0.00019$ & $0.01401 \pm 0.00011$ \\
\hline
\end{tabular}
\caption{Pre- and post-reconstruction errorbars on $\alpha_{\rm BAO}$ from Fisher forecast.
The mean of 10 single pre-reconstruction catalogs agrees with the sample covariance within a standard deviation.
For post-reconstruction, the difference is within 2 standard deviations.
}
\label{tab:sigma_alpha-pre-post}
\end{table}

\begin{figure}
\centering
\includegraphics[width=\columnwidth]{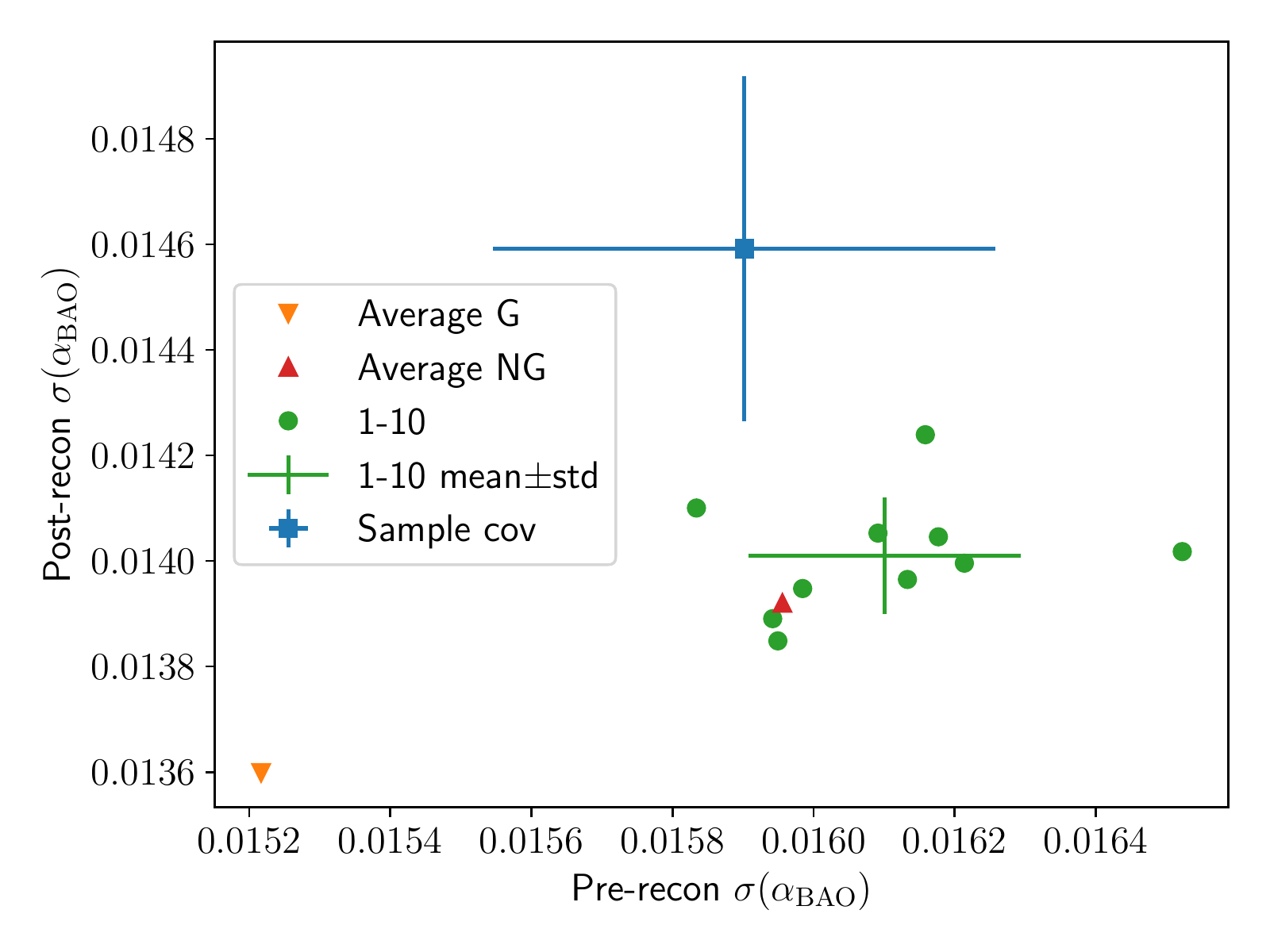}
\captionof{figure}{Pre- and post-reconstruction errorbars on $\alpha_{\rm BAO}$ from Fisher forecast plotted against each other.
Compared to the mock-averaged Gaussian run (Average G), the mock-averaged rescaled run (Average NG) and single-mock predictions with rescaling are closer to the sample covariance.
The horizontal (pre-reconstruction) agreement is closer than the vertical (post-reconstruction) one, but even the latter falls within $\approx 2\sigma$.
Note that the range of the axes is quite narrow, comprising only $\approx 9\%$ relative difference in errorbar before reconstruction and $\approx 7\%$ after.}
\label{fig:sigma_alpha-pre-post}
\end{figure}

For the sample covariance, we expect the variance of $C_{S,\alpha\alpha}$ to nearly follow Eq.~\ref{eq:X-covariance}:
\begin{eqnarray}
\Var \Args{C^{\rm par}_{S,\alpha\alpha}} \approx \frac2{n_S-1} \Arg{C^{\rm par}_{S,\alpha\alpha}}^2
\end{eqnarray}
and therefore the standard deviation of $\sigma(\alpha_{\rm BAO})$ of
\begin{eqnarray} \label{eq:sigma-sigma-alpha}
\sigma\Args{\sigma(\alpha_{\rm BAO})} \approx \frac{\sigma(\alpha_{\rm BAO})}{\sqrt{2(n_S-1)}},
\end{eqnarray}
resulting in relative precision of $2.2\%$.
This has been confirmed in Appendix~\ref{subsec:stats-validation}.

The resulting errorbar (Fisher) forecasts are provided in Table~\ref{tab:sigma_alpha-pre-post} and also presented as a scatter plot in Figure~\ref{fig:sigma_alpha-pre-post}.
We can notice that in any case the post-recon precision is expected to be higher than pre-recon.
In both pre-recon and post-recon, $\sigma(\alpha_{\rm BAO})$ in the mock-averaged clustering run without shot-noise rescaling are noticeably smaller than predicted from the sample covariance and are brought closer in the rescaled results.
Mock-averaged clustering with fit shot noise and single mock runs give very similar numbers.
The key conclusion is that single-mock runs are in good agreement with the sample covariance on $\sigma(\alpha_{\rm BAO})$, with a remarkably close match before reconstruction (just fractions of standard deviation) and a difference of $\approx 2$ standard deviation after reconstruction.
This gives assurance that data-based \rascalc{} covariances are on par with mock sample one for isotropic BAO fits.

\section{Summary and outlook}
\label{sec:conclusion}

This work continues a series of papers \citep{rascal,rascal-jackknife,rascalC,rascalC-legendre-3} developing a semi-empirical approach for estimating covariances of 2PCFs, combining analytical methods with the usage of measured clustering and calibration on jackknives.
The former brings smoothness and reliability, and the latter allows for flexibility of the results while being independent of mock galaxy catalogs.
We should note that the method is expected to be applicable at intermediate scales -- as analytical methods tend to fail on the smallest scales, while on the largest scales, the number of configurations increases making the computation longer, and the signal to noise in correlation function measurement decreases.
The latter issue could be alleviated by a smooth transition to a theoretically modeled 2PCF, which we leave for future work.

We have discussed the implications of split random-random counts computation and made a slight modification to the formalism to cover the reconstructed 2PCF estimates.
Then, we reconsidered the methods for covariance matrix comparison, paying great attention to their meaning, interpretation, and noise stemming from mock sample variance.

Finally, we have applied the selected approaches to the validation of \rascalc{} on single \themock{} catalogs (using their individual clustering measurements and shifted random catalogs after reconstruction), each representing a reasonable proxy for \desimtwo{} data, by comparison with full mock sample covariance.
We find a close agreement (maximum deviation $\approx 2.2\sigma$) with a perfect case, although much of this deviation is due to scatter in \rascalc{} results.
The preceding discussion about the interpretations of the metrics, focusing on a smaller number of observables and even fewer parameters allowed us to obtain a clearer quantitative assessment of the precision and accuracy of \rascalc{} results than in previous works.
One should keep in mind the mocks are approximate and this can partially account for the imperfection of the match with the reference statistics.

Focusing on the errorbar of the BAO scale, we found a very close, percent-level agreement with the sample covariance from mocks.
It is on par with the accuracy that a set of $\approx 1000$ simulations can provide.
The number of available mocks thus limits the precision of the validation at the current level.

The comparison suggests that noise in the input 2PCF might be a significant limiting factor for the accuracy of our covariance matrices in higher-dimensional spaces.
Smoothing this input or complementing it with a theoretical best-fit model could help to mitigate this issue without introducing many additional assumptions.
This marks an important topic for follow-up studies.

In the full measurement space before reconstruction, using a shot-noise rescaling is particularly clearly beneficial compared to the pure Gaussian estimate even with less noisy (mock sample average) input clustering.
The discrepancies and fluctuations are likely to be impacted by the precision of correlation function estimates from data, which will improve with its size in the future.
Further validations with larger mocks, corresponding to a year and/or full five years of DESI data, will follow.

During the comparison, we have seen indications that the reconstructed extension may not be working better than with the pre-reconstructed data, contrary to the expectation.
This might be related to the subtleties of small-scale behavior of reconstructed points and will be investigated in further detail in future work.

Another perspective direction is the development of alternatives to shot-noise rescaling within the more generic semi-analytic configuration-space formalism.
Usage of fully empirical higher-point functions is likely to be not viable, due to a significantly higher number of bins and accordingly lower signal-to-noise.
Precise theoretical modeling of non-Gaussian correlation functions is also very challenging.
Instead, we might include a basic prescription for non-Gaussian covariance contribution inferred from a set of detailed simulations, or use approximate expressions for higher-point functions like $\zeta(\vec r_1,\vec r_2) = Q\big[\xi(r_1)\xi(r_2)+\xi(r_1)\xi(|\vec r_1-\vec r_2|) + \xi(r_2)\xi(|\vec r_1-\vec r_2|)\big]$ motivated by hierarchical models \citep{hierarchical-3PCF}, and possibly a similar structure for the 4PCF.
This might provide better accuracy than rescaling the Gaussian terms while keeping the number of parameters low and thus still allowing us to fit them to a reference (e.g. jackknife) covariance.
On the other hand, the aforementioned 3PCF prescription is known to be far from exact with constant $Q$ \citep{Q-3PCF}, and the computations may suffer from slower convergence due to additional large values of small-scale 2PCF compared to Gaussian parts.

The key advantage of the approaches considered in this paper is that a covariance can be based on the data itself and does not require matching mock catalogs each time.
This alleviates the concern about the accuracy of such approximate simulations, not completely removing it since we still need some as references for validation of the prescriptions.
Perhaps more importantly, covariance computation becomes much more flexible.
Relevant cases are when the clustering signal changes after more data are gathered, or when alternative assumptions (for instance, the base cosmology) are tested.
For such updates, calibration, generation, and processing of a suite of mocks large enough for good sample covariance consumes great resources.
The computation of one proxy-data covariance in this paper took about 100 core-hours for pre-recon and about 300 core-hours for post-recon\footnote{In one case out of ten (both pre- and post-recon), a repeated computation taking twice longer was performed.}.
This could be optimized further by noting that the intrinsic consistency in each run was much higher than the resemblance of reference sample covariance -- time of computation helps with the former but the latter is fundamentally limited by the accuracy of approximations and precision of the input correlation function.
Moreover, the number of mocks limits the level of accuracy of validation.
Fast covariance matrix computation without mocks can also allow shifting the balance from quantity to quality of the simulations, freeing resources for more detailed ones.

\section*{Acknowledgements}

We thank Daniel Forero-Sanchez, Violeta Gonzalez-Perez, Nikhil Padmanabhan, Will Percival, Oliver Philcox, and Martin White for useful comments and fruitful discussions.
MR and DJE are supported by the U.S. Department of Energy (DOE) grant DE-SC0007881 and by the Simons Foundation Investigator program.

This material is based upon work supported by the U.S. Department of Energy (DOE), Office of Science, Office of High-Energy Physics, under Contract No. DE–AC02–05CH11231, and by the National Energy Research Scientific Computing Center, a DOE Office of Science User Facility under the same contract. Additional support for DESI was provided by the U.S. National Science Foundation (NSF), Division of Astronomical Sciences under Contract No. AST-0950945 to the NSF’s National Optical-Infrared Astronomy Research Laboratory; the Science and Technology Facilities Council of the United Kingdom; the Gordon and Betty Moore Foundation; the Heising-Simons Foundation; the French Alternative Energies and Atomic Energy Commission (CEA); the National Council of Science and Technology of Mexico (CONACYT); the Ministry of Science and Innovation of Spain (MICINN), and by the DESI Member Institutions: \url{https://www.desi.lbl.gov/collaborating-institutions}. Any opinions, findings, and conclusions or recommendations expressed in this material are those of the author(s) and do not necessarily reflect the views of the U. S. National Science Foundation, the U. S. Department of Energy, or any of the listed funding agencies.

The authors are honored to be permitted to conduct scientific research on Iolkam Du’ag (Kitt Peak), a mountain with particular significance to the Tohono O’odham Nation.

\section*{Data availability}

The code is openly accessible at \url{https://github.com/oliverphilcox/RascalC}.
All data from the tables and figures are available in machine-readable format at \doi{10.5281/zenodo.7750637} in compliance with the DESI data management plan.
The \themocks{} used in this paper will be made public with the DESI Y1 data release (DR1), and all the covariance matrices used in this work will be released in the same supplementary material \citep{supp-mat}.

\bibliographystyle{mnras}
\bibliography{references}



\appendix

\onecolumn

\section{Covariance estimators}
\label{sec:previous-work-ext}

\subsection{Full covariance in radial and angular bins}
\label{subsec:full-cov-est}

The expression for full covariance in radial and angular bins ($\cov \Args{\Arg{\hat\xi^{XY}}_a^c, \Arg{\hat\xi^{ZW}}_b^d}$) is
\begin{eqnarray} \label{eq:Cov2x2estimator}
\Arg{\tilde C^{XY,ZW}}_{ab}^{cd} \Arg{\bm \alpha_{\rm SN}} &=& \Arg{^4C^{XY,ZW}}_{ab}^{cd} + \frac{\alpha_{\rm SN}^X}4 \Args{\delta^{XW} \Arg{^3C^{X,YZ}}_{ab}^{cd} + \delta^{XZ} \Arg{^3C^{X,YW}}_{ab}^{cd}} \\ \nonumber
&+& \frac{\alpha_{\rm SN}^Y}4 \Args{\delta^{YW} \Arg{^3C^{Y,XZ}}_{ab}^{cd} + \delta^{YZ} \Arg{^3C^{Y,XW}}_{ab}^{cd}}
+ \frac{\alpha_{\rm SN}^X \alpha_{\rm SN}^Y}2 \Args{\delta^{XW} \delta^{YZ} + \delta^{XZ} \delta^{YW}} \Arg{^2C^{XY}}_{ab}^{cd}
\end{eqnarray}
with
\begin{eqnarray} \label{eq:Cov2x2_234_Point_Defs}
\Arg{^4C^{XY,ZW}}_{ab}^{cd} &=& \frac{1}{\left(R^XR^Y\right)_a^c \left(R^ZR^W\right)_b^d} \sum_{i\neq j\neq k\neq l} n^X_i n^Y_j n^Z_k n^W_l w^X_i w^Y_j w^Z_k w^W_l \Theta^a(r_{ij}) \Theta^c(\mu_{ij}) \Theta^b(r_{kl}) \Theta^d(\mu_{kl}) \\ \nonumber
& & \times \Args{\eta^{({\rm c}),XYWZ}_{ijkl} + 2\xi^{XZ}_{ik} \xi^{YW}_{jl}} \\ \nonumber
\Arg{^3C^{Y,XZ}}_{ab}^{cd} &=& \frac{4}{\left(R^XR^Y\right)_a^c \left(R^YR^Z\right)_b^d} \sum_{i\neq j\neq k} n^X_i n^Y_j n^Z_k w^X_i \Arg{w^Y_j}^2 w^Z_k \Theta^a(r_{ij}) \Theta^c(\mu_{ij}) \Theta^b(r_{jk}) \Theta^d(\mu_{jk}) \Args{\zeta^{XYZ}_{ijk}+\xi^{XZ}_{ik}} \\ \nonumber
\Arg{^2C^{XY}}_{ab}^{cd} &=& \frac{2\delta^{ab}\delta^{cd}}{\left(R^XR^Y\right)_a^c \left(R^XR^Y\right)_b^d}\sum_{i\neq j} n^X_i n^Y_j \Arg{w^X_i w^Y_j}^2 \Theta^a(r_{ij}) \Theta^c(\mu_{ij}) \Args{1+\xi^{XY}_{ij}}
\end{eqnarray}
where $\delta^{XY}$, $\delta^{ab}$ and $\delta^{cd}$ are Kronecker deltas.
Similarly to \citet{rascalC}, we have written $2\xi_{ik}\xi_{jl}$ instead of $\xi_{ik}\xi_{jl}+\xi_{il}\xi_{jk}$, allowing to optimize the computation using the symmetries of the term, but requiring to compute a few more distinct terms in multi-tracer setup.
Computing full sums (beyond the pair one) is not feasible, so are estimated by the Monte-Carlo method instead, by randomly sampling a subset of 2/3/4-point configurations from the random catalog \citep{rascalC}.

The most practical form for single tracer $X$ is simpler:
\begin{eqnarray} \label{eq:Cov2x2estimator1}
\Arg{C^{XX,XX}}_{ab}^{cd} \Arg{\alpha_{\rm SN}^X} = \Arg{^4C^{XX,XX}}_{ab}^{cd} + \alpha_{\rm SN}^X \Arg{^3C^{X,XX}}_{ab}^{cd} + \Arg{\alpha_{\rm SN}^X}^2 \Arg{^2C^{XX}}_{ab}^{cd}.
\end{eqnarray}

\subsection{Jackknife covariance in radial and angular bins}
\label{subsec:jack-cov-est}

The jackknife pair counts are
\begin{eqnarray} \label{eq:NN-RR-discrete-jack}
\Arg{N^XN^Y_A}_a^c &=& \sum_{i\neq j}n^X_in^Y_jw^X_iw^Y_j\Theta^a(r_{ij})\Theta^c(\mu_{ij})\delta^X_i\delta^Y_j q^A_{ij} \\ \nonumber
\Arg{R^XR^Y_A}_a^c &=& \sum_{i\neq j}n^X_in^Y_jw^X_iw^Y_j\Theta^a(r_{ij})\Theta^c(\mu_{ij}) q^A_{ij},
\end{eqnarray}
where $q^A_{ij}$ is the jackknife weighting factor for the pair of particles, and for any pair $\sum_A q^A_{ij} = 1$ in the unrestricted jackknife formalism.

The binned correlation function estimate $\Arg{\hat\xi^{XY}_A}_a^c$ is their ratio.
It is sensible to weight the regions by the $RR$ pair counts, which roughly correspond to the volume fraction of each region:
\begin{eqnarray} \label{eq:jack-weights}
\Arg{w^{XY}_A}_a^c = \frac{\Arg{R^XR^Y_A}_a^c}{\Arg{R^XR^Y}_a^c},
\end{eqnarray}
and then the weighted average of the jackknife 2PCF is identical to full-survey estimate (Eq.~\eqref{eq:Landy-Szalay-binned}).

The jackknife covariance estimate is
\begin{eqnarray} \label{eq:cov-jackknife-def}
\Arg{C_J^{XY,ZW}}_{ab}^{cd} = \frac1{1 - \sum_B \Arg{w^{XY}_B}_a^c \Arg{w^{ZW}_B}_b^d} \sum_A \Argf{\Arg{w^{XY}_A}_a^c \Arg{w^{ZW}_A}_b^d \Args{\Arg{\hat\xi^{XY}_A}_a^c - \Arg{\hat\xi^{XY}}_a^c} \Args{\Arg{\hat\xi^{ZW}_A}_b^d - \Arg{\hat\xi^{ZW}}_b^d}}.
\end{eqnarray}
By substituting Eq.~\eqref{eq:Landy-Szalay-binned}, expanding using Eqs.~\eqref{eq:NN-RR-discrete-jack}~\&~\eqref{eq:NN-RR-discrete} and simplifying through Eq.~\eqref{eq:shot-noise-approximation} one can arrive to
\begin{eqnarray} \label{eq:Cov2x2estimator-jack}
\Arg{\tilde C_J^{XY,ZW}}_{ab}^{cd} = \Arg{^4C_J^{XY,ZW}}_{ab}^{cd} + \frac{\alpha^X}4 \Args{\delta^{XW} \Arg{^3C_J^{X,YZ}}_{ab}^{cd} + \delta^{XZ} \Arg{^3C_J^{X,YW}}_{ab}^{cd}} + \frac{\alpha^Y}4 \Args{\delta^{YW} \Arg{^3C_J^{Y,XZ}}_{ab}^{cd} + \delta^{YZ} \Arg{^3C_J^{Y,XW}}_{ab}^{cd}} \\ \nonumber
+ \frac{\alpha^X \alpha^Y}2 \Args{\delta^{XW} \delta^{YZ} + \delta^{XZ} \delta^{YW}} \Arg{^2C_J^{XY}}_{ab}^{cd}
\end{eqnarray}
with
\begin{eqnarray} \label{eq:Cov2x2_234_Point_Defs-jack}
\Arg{^4C_J^{XY,ZW}}_{ab}^{cd} &=& \frac{1}{\left(R^XR^Y\right)_a^c \left(R^ZR^W\right)_b^d \Args{1 - \sum_B \Arg{w^{XY}_B}_a^c \Arg{w^{ZW}_B}_b^d}} \sum_{i\neq j\neq k\neq l} n^X_i n^Y_j n^Z_k n^W_l w^X_i w^Y_j w^Z_k w^W_l \Theta^a(r_{ij}) \Theta^c(\mu_{ij}) \\ \nonumber
& & \times \Theta^b(r_{kl}) \Theta^d(\mu_{kl}) \Args{\eta^{({\rm c}),XYWZ}_{ijkl} + \xi^{XY}_{ij} \xi^{ZW}_{kl} + 2\xi^{XZ}_{ik} \xi^{YW}_{jl}} \Arg{\omega^{XY,ZW}_{ijkl}}_{ab}^{cd} \\ \nonumber
\Arg{^3C_J^{Y,XZ}}_{ab}^{cd} &=& \frac{4}{\left(R^XR^Y\right)_a^c \left(R^YR^Z\right)_b^d \Args{1 - \sum_B \Arg{w^{XY}_B}_a^c \Arg{w^{YZ}_B}_b^d}} \sum_{i\neq j\neq k} n^X_i n^Y_j n^Z_k w^X_i \Arg{w^Y_j}^2 w^Z_k \Theta^a(r_{ij}) \Theta^c(\mu_{ij}) \Theta^b(r_{jk}) \Theta^d(\mu_{jk}) \\ \nonumber
& & \times \Args{\zeta^{XYZ}_{ijk}+\xi^{XZ}_{ik}} \Arg{\omega^{XY,YZ}_{ijjk}}_{ab}^{cd} \\ \nonumber
\Arg{^2C_J^{XY}}_{ab}^{cd} &=& \frac{2\delta^{ab}\delta^{cd}}{\left(R^XR^Y\right)_a^c \left(R^XR^Y\right)_b^d \Args{1 - \sum_B \Arg{w^{XY}_B}_a^c \Arg{w^{XY}_B}_b^d}}\sum_{i\neq j} n^X_i n^Y_j \Arg{w^X_i w^Y_j}^2 \Theta^a(r_{ij}) \Theta^c(\mu_{ij}) \Args{1+\xi^{XY}_{ij}} \Arg{\omega^{XY,XY}_{ijij}}_{ab}^{cd},
\end{eqnarray}
where $\Arg{\omega_{ijkl}}_{ab}^{cd}$ is an additional weight tensor:
\begin{eqnarray}
\Arg{\omega^{XY,ZW}_{ijkl}}_{ab}^{cd} = \sum_A \Argf{\Args{q^A_{ij} - \Arg{w^{XY}_A}_a^c} \Args{q^A_{kl} - \Arg{w^{ZW}_A}_b^d}}.
\end{eqnarray}
In practice, shot-noise rescaling for each tracer has been obtained by fitting the prediction for jackknife covariance of its auto-correlation function
\begin{eqnarray} \label{eq:Cov2x2estimator1-jack}
\Arg{C_J^{XX,XX}}_{ab}^{cd} = \Arg{^4C_J^{XX,XX}}_{ab}^{cd} + \alpha^X \Arg{^3C_J^{X,XX}}_{ab}^{cd} + \Arg{\alpha^X}^2 \Arg{^2C_J^{XX}}_{ab}^{cd}
\end{eqnarray}
to the data-based jackknife estimate (computed via Eq.~\eqref{eq:cov-jackknife-def}).
The resulting $\alpha^X$ value(s) should be plugged into Eq.~\eqref{eq:Cov2x2estimator} to obtain the full survey covariance.

\subsection{Covariance of Legendre multipoles in radial bins}
\label{subsec:Legendre-cov-est}

\cite{rascalC-legendre-3} further derive direct covariance estimators for Legendre moments of the anisotropic 2PCF, which is related to $\xi(r,\mu)$ through
\begin{eqnarray}
\xi^{XY}(r,\mu) &=& \sum_{\ell=0}^{\infty} \Arg{\xi^{XY}}^\ell(r)L_\ell(\mu), \\
\label{eq:Legendre-from-angular-2PCF}
\Arg{\xi^{XY}}^\ell(r) &=& \frac{2\ell+1}{2}\int_{-1}^1d\mu\, \xi^{XY}(r,\mu)L_\ell(\mu) =\frac{(-1)^\ell+1}{2}(2\ell+1)\int_0^1d\mu\, \xi^{XY}(r,\mu)L_\ell(\mu),
\end{eqnarray}
$L_\ell(\mu)$ being the Legendre polynomial of order $\ell$, and the second equality in last line assumes symmetry $\xi^{XY}(r,\mu) = \xi^{XY}(r,-\mu)$ (necessarily true for auto-correlation, $X=Y$, and violation in cross-correlations is debatable).

While one could estimate the angularly binned 2PCF first and then transform it to Legendre moments, a direct computation of the latter allows to evade inaccuracies caused by the replacement of integral by sum over bins, and removes the need for very fine splitting in $\mu$, which would make the covariance integrals slower to converge.

Averaging Eq.~\eqref{eq:Legendre-from-angular-2PCF} between the boundaries of radial bin $a$, we obtain
\begin{eqnarray} \label{eq:Legendre-from-angular-2PCF-v2}
\Arg{\xi^{XY}}^\ell_a = (2\ell+1)\int_0^1d\mu\,\Arg{\xi^{XY}}_a(\mu)L_\ell(\mu).
\end{eqnarray}
Here $\xi^{XY}_a(\mu)$, binned radially but not angularly, is given by the reduced Landy-Szalay estimator
\begin{eqnarray}
\Arg{\xi^{XY}}_a(\mu) = \frac{\Arg{N^XN^Y}_a(\mu)}{\Arg{R^XR^Y}_a(\mu)}
\end{eqnarray}
similarly to Eqs.~\eqref{eq:Landy-Szalay}~\&~\eqref{eq:Landy-Szalay-binned}.
The continuous counts $PP_a(\mu)$ (where each $P$ can stand for $D^X$, $D^Y$, $R^X$ or $R^Y$) are the limit of the ratio of conventional pair counts in infinitesimally small angular bins $c$ to its width $\delta\mu$,
$PP_a^c = \int d\mu\,PP_a(\mu)\Theta^c(\mu) \approx PP_a(\mu_c)\delta \mu$,
$\mu_c$ being the bin center.

A functional form for the continuous random pair counts $\Arg{R^XR^Y}_a\Arg{\mu}$ is needed to go further due to integration in Eq.~\eqref{eq:Legendre-from-angular-2PCF-v2}.
This is done via a {\it survey correction function} $\Phi^{XY}(r_a,\mu)$ (following \cite{survey-correction}), which accounts for survey boundaries and selection, defined through
\begin{eqnarray} \label{eq:RR_approximation_generalized}
\left(R^{X}R^{Y}\right)_a(\mu) \equiv \frac{V^X \bar n^X \bar n^Y \bar w^X \bar w^Y v_a}{\Phi^{XY}(r_a,\mu)},
\end{eqnarray}
where $V^X$ is the total volume of survey for tracer $X$, and $\bar n^X$ and $\bar w^X$ are the survey-averaged number density and weight for tracer $X$, respectively.
$v_a = \tfrac{4\pi}{3}\left(r^3_{a,\mathrm{max}}-r^3_{a,\mathrm{min}}\right)$ is the volume of radial bin $a$ ($r_{a,\min}$ and $r_{a,\max}$ being its lower and upper boundaries).
The numerator is the expression for a periodic box with uniform number density, weights and bins in $\Abs{\mu}$ (i.e. negative values reversed into $[0,1]$ interval), thus in this simple case the survey correction function $\Phi^{XY}(r_a,\mu)=1$.
For nontrivial geometry, varying density and/or weights, it may become different but not too far by order of magnitude.
However, note that only the angular variable value is arbitrary, while the radius is taken as representative of the bin.
In practice, a piecewise-polynomial functional fit to empirical data was used by \cite{rascalC-legendre-3}.

Then the 2PCF Legendre moment can be estimated
and the covariance can be computed by definition, using Eq.~\eqref{eq:shot-noise-approximation}, resulting in
\begin{eqnarray} \label{eq:cov2x2_Legendre_generalized}
\left(C^{XY,ZW}\right)_{ab}^{pq} &=& \left({}^4C^{XY,ZW}\right)_{ab}^{pq} + \frac{\alpha^X}{4} \left[\delta^{XW}\left({}^3C^{X,YZ}\right)_{ab}^{pq}+\delta^{XZ}\left({}^3C^{X,YW}\right)_{ab}^{pq}\right] \\ \nonumber
&+&\frac{\alpha^Y}{4}\left[\delta^{YW}\left({}^3C^{Y,XZ}\right)_{ab}^{pq}+\delta^{YZ}\left({}^3C^{Y,XW}\right)_{ab}^{pq} \right]
\frac{\alpha^X\alpha^Y}{2}\left(\delta^{XW}\delta^{YZ}+\delta^{XZ}\delta^{YW}\right)\left({}^2C^{XY}\right)_{ab}^{pq}.
\end{eqnarray}
\begin{eqnarray} \label{eq:cov2x2_Legendre_terms_generalized}
    \left({}^4C^{XY,ZW}\right)_{ab}^{pq} &=& \frac{(2p+1)(2q+1)}{V^X V^Z \bar n^X \bar n^Y \bar n^Z \bar n^W \bar w^X \bar w^Y \bar w^Z \bar w^W v_a v_b}\sum_{i\neq j\neq k \neq l}n^X_in^Y_jn^Z_kn^W_lw^X_iw^Y_jw^Z_kw^W_l\Theta^a(r_{ij})\Theta^b(r_{kl})\\\nonumber
    &\times&\Phi^{XY}(r_a,\mu_{ij})\Phi^{ZW}(r_b,\mu_{kl})L_p(\mu_{ij})L_q(\mu_{kl})\left[\xi^{(4),XYZW}_{ijkl}+\xi^{XZ}_{ik}\xi^{YW}_{jl}+\xi^{XW}_{il}\xi^{YZ}_{jk}\right]\\\nonumber
    \left({}^3C^{Y,XZ}\right)_{ab}^{pq} &=& 4\times \frac{(2p+1)(2q+1)}{V^X V^Z \bar n^X \Arg{\bar n^Y}^2 n^Z \bar w^X \Arg{\bar w^Y}^2 w^Z v_a v_b}\sum_{i\neq j\neq k}n^X_in^Y_jn^Z_kw^X_i\left(w^Y_j\right)^2w^Z_k\Theta^a(r_{ij})\Theta^b(r_{jk})\\\nonumber
    &\times&\Phi^{XY}(r_a,\mu_{ij})\Phi^{ZY}(r_b,\mu_{jk})L_p(\mu_{ij})L_q(\mu_{jk})\left[\zeta^{XYZ}_{ijk}+\xi^{XZ}_{ik}\right]\\\nonumber
    \left({}^2C^{XY}\right)_{ab}^{pq} &=& 2\delta_{ab}\times \frac{(2p+1)(2q+1)}{\left(V^X \bar n^X \bar n^Y \bar w^X \bar w^Y v_a\right)^2}\sum_{i\neq j}n^X_in^Y_j\left(w^X_iw^Y_j\right)^2\Theta^a(r_{ij})\\\nonumber
    &\times&\left(\Phi^{XY}(r_a,\mu_{kl})\right)^2L_p(\mu_{ij})L_q(\mu_{ij})\left[1+\xi^{XY}_{ij}\right].
\end{eqnarray}

Jackknife poses certain challenges to direct computation for the Legendre moments of the 2PCF.
So far it has been suggested that the shot-noise rescaling value(s) should be optimized with a jackknife estimate on angularly binned 2PCF \citep{rascalC-legendre-3}.

\section{Statistics of comparison metrics for noisy sample covariance matrix}
\label{sec:cov-comparison-properties}

Here we provide derivations of the expectation values for comparison metrics between a noisy sample covariance and the true covariance/precision matrix.
This is useful for testing how close \rascalc{} results are to the latter.

\subsection{KL divergence mean and variance}
\label{subsec:kl-div-stats}

A more generic setup -- two sample covariance matrices based on draws from a multivariate normal distribution -- has been considered in Appendix~D of \cite{rascalC}.
However, the derivation was limited to the expectation value of the KL divergence between them, and we have not been able to find a reference about the metric's scatter around the mean (variance or standard deviation).
In addition, we believe the final result there is slightly incorrect, namely $1$ should be subtracted from the number of samples.
This is because for the estimate of sample covariance commonly used with mocks
\begin{eqnarray}
X_{ab} = \frac1{n_S-1} \sum_{i=1}^{n_S} (x_{a,i} - \bar x_a) (x_{b,i} - \bar x_b)
\end{eqnarray}
the mean is not known beforehand but estimated from the sample as well: $\bar x_a \equiv \frac1{n_S} \sum_{i=1}^{n_S} x_{a,i}$.
This reduces the number of degrees of freedom by one.
Then the covariance of sample covariance matrix elements is
\begin{eqnarray} \label{eq:X-covariance}
\cov (X_{ab}, X_{cd}) = \frac{C_{0,ac} C_{0,bd} + C_{0,ad} C_{0,bc}}{n_S-1}
\end{eqnarray}
instead of
$$ \cov (X_{ab}, X_{cd}) = \frac{C_{0,ac} C_{0,bd} + C_{0,ad} C_{0,bc}}{n_S} $$
as in \cite{rascalC}.
${\bm C}_0$ is the true underlying covariance matrix of the Gaussian distribution the samples are drawn from.
The sample covariance estimate is unbiased, meaning that the expectation value is the true covariance: $\Avg{\bm X}={\bm C}_0$.
Then, considering two sample covariance matrices ${\bf X}_i$ obtained from $n_S^{(i)}$ samples each, decomposing them as ${\bf X}_i={\bf C}_0+{\bf \delta X}_i$, Taylor expanding and only leaving the leading nontrivial (quadratic) order in ${\bf \delta X}_i$, we obtain
\begin{eqnarray} \label{eq:D_KL-expectation-2samples}
\Avg{D_{\rm KL} \Arg{{\bf X}_1^{-1}, {\bf X}_2}} \approx \frac{N_{\rm bins}(N_{\rm bins}+1)}4 \Arg{\frac1{n_S^{(1)}-1} + \frac1{n_S^{(2)}-1}}
\end{eqnarray}
instead of
$$ \Avg{D_{\rm KL} \Arg{{\bf X}_1^{-1}, {\bf X}_2}} \approx \frac{N_{\rm bins}(N_{\rm bins}+1)}4 \Arg{\frac1{n_S^{(1)}} + \frac1{n_S^{(2)}}} $$
as in \cite{rascalC}.

Since in this work we only consider one sample covariance matrix, while the \rascalc{} results are not expected to follow the Wishart distribution, the more relevant result is for the true precision matrix ${\bf \Psi}_0$:
\begin{eqnarray} \label{eq:D_KL-expectation}
\Avg{D_{\rm KL} \Arg{{\bf \Psi}_0, {\bf X}}} \approx \frac{N_{\rm bins}(N_{\rm bins}+1)}{4(n_S-1)},
\end{eqnarray}
which can be obtained from Eq.~\eqref{eq:D_KL-expectation-2samples} by setting the first number of samples to infinity, reducing the noise in ${\bm X}_1^{-1}$ to zero.

For the further derivations, it is convenient to ``normalize'' the covariance. Let us take
\begin{eqnarray} \label{eq:measurement-normalization}
\bm y_i = {\bf \Psi}_0^{1/2} \bm x_i,
\end{eqnarray}
where ${\bf \Psi}_0^{1/2}$ means the matrix square root of ${\bf \Psi}_0$ -- a matrix with the same eigenvectors and eigenvalues equal to the square roots of corresponding eigenvalues of the original matrix.
Then
\begin{eqnarray}
\cov (y_{a,i}, y_{b,j}) = \delta^{ij} \delta^{ab}.
\end{eqnarray}
Let us also introduce $\bar y_a \equiv \frac1{n_S} \sum_{i=1}^{n_S} y_{a,i}$, define $\delta y_{a,i} \equiv y_{a,i} - \bar y_a$ (so that $\left\langle \delta y_{a,i} \right\rangle = 0$) and finally compute the ``normalized'' covariance matrix:
\begin{eqnarray} \label{eq:Y-definition}
Y_{ab} \equiv \frac1{n_S-1} \sum_{i=1}^{n_S} \delta y_{a,i} \delta y_{b,i}.
\end{eqnarray}
Then also
\begin{eqnarray} \label{eq:Y-relation}
{\bf Y} = {\bf \Psi}_0^{1/2} {\bf X \Psi}_0^{1/2}.
\end{eqnarray}
Let us compute
\begin{eqnarray}
\cov (\delta y_{a,i}, \delta y_{b,j}) = \left\langle \delta y_{a,i} \delta y_{b,j} \right\rangle = \delta^{ij} \delta^{ab} - 2 \times \frac1{n_S} \delta^{ab} + n_S \times \frac1{n_S^2} \delta^{ab} = \left( \delta^{ij} - \frac1{n_S} \right) \delta^{ab}.
\end{eqnarray}
As a consequence, $\Avg{\bf Y} = \bf \mathbb I$, and we can expand
\begin{eqnarray} \label{eq:delta-Y}
{\bf Y} = {\bf \mathbb I} + {\bf \delta Y}
\end{eqnarray}
while $\Avg{\bf \delta Y} = 0$. Also,
\begin{eqnarray} \label{eq:Y-covariance}
\cov (Y_{ab}, Y_{cd}) = \left\langle \delta Y_{ab} \delta Y_{cd} \right\rangle = \frac{\delta^{ac} \delta^{bd} + \delta^{ad} \delta^{bc}}{n_S-1}.
\end{eqnarray}
Now let us expand the KL divergence using the ``normalized'' covariance matrix, starting from
\begin{eqnarray}
2 D_{\rm KL} \Arg{{\bf \Psi}_0, {\bf X}} = \tr \Arg{{\bf \Psi}_0 {\bf X}} - N_{\rm bins} - \ln \det \Arg{{\bf \Psi}_0 {\bf X}},
\end{eqnarray}
we can write ${\bf \Psi}_0 = {\bf \Psi}_0^{1/2} {\bf \Psi}_0^{1/2}$, use the cyclic property of trace and determinant to arrive to
\begin{eqnarray}
2 D_{\rm KL} \Arg{{\bf \Psi}_0, {\bf X}} = \tr \Arg{\bf Y} - N_{\rm bins} - \ln \det \Arg{\bf Y},
\end{eqnarray}
remembering Eq.~\eqref{eq:Y-relation}.
Then we expand in $\bf \delta Y$ (Eq.~\eqref{eq:delta-Y}):
\begin{eqnarray}
2 D_{\rm KL} \Arg{{\bf \Psi}_0, {\bf X}} = \tr \Arg{\bf \delta Y} - \ln \det \Arg{\bf \mathbb I + \delta Y}.
\end{eqnarray}
Now using $\ln \det {\bf A} = \tr \ln {\bf A}$ and expanding the second term in Taylor series up to quadratic order in $\bf \delta Y$ we obtain
\begin{eqnarray} \label{eq:D_KL-Y-approx}
2 D_{\rm KL} \Arg{{\bf \Psi}_0, {\bf X}} \approx \frac12 \tr \Args{\Arg{\bf \delta Y}^2} = \frac12 \sum_{a,b=1}^{N_{\rm bins}} \delta Y_{ab} \delta Y_{ab}.
\end{eqnarray}
Taking the expectation value of Eq.~\eqref{eq:D_KL-Y-approx} and using Eq.~\eqref{eq:Y-covariance}, one can re-derive Eq.~\eqref{eq:D_KL-expectation}. We will proceed to compute the variance:
\begin{eqnarray}
\Var \Argf{\tr \Args{\Arg{\bf \delta Y}^2}} = \Avg{\Arg{\tr \Args{\Arg{\bf \delta Y}^2}}^2} - \Avg{\tr \Args{\Arg{\bf \delta Y}^2}}^2 = \sum_{a,b,c,d=1}^{N_{\rm bins}} [\left\langle \delta Y_{ab} \delta Y_{ab} \delta Y_{cd} \delta Y_{cd} \right\rangle - \left\langle \delta Y_{ab} \delta Y_{ab} \right\rangle \left\langle \delta Y_{cd} \delta Y_{cd} \right\rangle].
\end{eqnarray}
Full expansion gives
\begin{eqnarray}
Y_{ab} Y_{ab} Y_{cd} Y_{cd} = \frac1{(n_S-1)^4} \sum_{i,j,k,l=1}^{n_S} \delta y_{a,i} \delta y_{b,i} \delta y_{a,j} \delta y_{b,j} \delta y_{c,k} \delta y_{d,k} \delta y_{c,l} \delta y_{d,l}.
\end{eqnarray}
$\delta y$ are normally distributed and have zero means, so for them, we can use Wick's theorem to split this into all possible pairs. ($\bf \delta Y$ also has zero mean, but not Gaussian distribution, this is why we need to go to a deeper level.) The total number of pairs is $8!/(4!\times 2^4)=105$, so it is easy to go over them in a computer program. Additionally, it is useful to check which are similar. It is apparent that the following five index permutations leave the expression unchanged: $a \leftrightarrow b$, $c \leftrightarrow d$, $i \leftrightarrow j$, $k \leftrightarrow l$ and $(a,b,i,j) \leftrightarrow (c,d,k,l)$. Finally, some of the pairs will not contribute to variance and can be excluded: contraction of a pair inside the same $Y$ contribute to $\left\langle Y \right\rangle = \mathbb I$ and must be subtracted; and if all the contracted pairs correspond to $Y$'s with the same indices, that contributes to the mean of $D_{\rm KL}$ and has to be subtracted too.

We find there are 56 pair assignments contributing to $\approx 16(n_S-1)^4 \Var [D_{\rm KL} (\Psi_0, X)]$, but no more than 8 of them are distinct after using symmetries:
\begin{eqnarray}
& 8 \sum_{a,b,c,d=1}^{N_{\rm bins}} \sum_{i,j,k,l=1}^{n_S} \left\langle \delta y_{a,i} \delta y_{a,j} \right\rangle \left\langle \delta y_{b,i} \delta y_{c,k} \right\rangle \left\langle \delta y_{b,j} \delta y_{c,l} \right\rangle \left\langle \delta y_{d,k} \delta y_{d,l} \right\rangle = \nonumber \\
& 8 \sum_{a,b,c,d=1}^{N_{\rm bins}} \sum_{i,j,k,l=1}^{n_S} \left( \delta^{ij} - \frac1{n_S} \right) \left( \delta^{ik} - \frac1{n_S} \right) \delta^{bc} \left( \delta^{jl} - \frac1{n_S} \right) \delta^{bc} \left( \delta^{kl} - \frac1{n_S} \right)
= 8 N_{\rm bins}^3 (n_S - 1) \\
& 16 \sum_{a,b,c,d=1}^{N_{\rm bins}} \sum_{i,j,k,l=1}^{n_S} \left\langle \delta y_{a,i} \delta y_{a,j} \right\rangle \left\langle \delta y_{b,i} \delta y_{c,k} \right\rangle \left\langle \delta y_{b,j} \delta y_{d,l} \right\rangle \left\langle \delta y_{d,k} \delta y_{c,l} \right\rangle = \nonumber \\
& 16 \sum_{a,b,c,d=1}^{N_{\rm bins}} \sum_{i,j,k,l=1}^{n_S} \left( \delta^{ij} - \frac1{n_S} \right) \left( \delta^{ik} - \frac1{n_S} \right) \delta^{bc} \left( \delta^{jl} - \frac1{n_S} \right) \delta^{bd} \left( \delta^{kl} - \frac1{n_S} \right) \delta^{dc} = 16 N_{\rm bins}^2 (n_S-1) \\
& 8 \sum_{a,b,c,d=1}^{N_{\rm bins}} \sum_{i,j,k,l=1}^{n_S} \left\langle \delta y_{a,i} \delta y_{b,j}\right\rangle \left\langle \delta y_{b,i} \delta y_{c,k} \right\rangle \left\langle \delta y_{a,j} \delta y_{d,l} \right\rangle \left\langle \delta y_{d,k} \delta y_{c,l} \right\rangle = \nonumber \\
& 8 \sum_{a,b,c,d=1}^{N_{\rm bins}} \sum_{i,j,k,l=1}^{n_S} \left( \delta^{ij} - \frac1{n_S} \right) \delta^{ab} \left( \delta^{ik} - \frac1{n_S} \right) \delta^{bc} \left( \delta^{jl} - \frac1{n_S} \right) \delta^{ad} \left( \delta^{kl} - \frac1{n_S} \right) \delta^{cd} = 8 N_{\rm bins} (n_S-1) \\
& 4 \sum_{a,b,c,d=1}^{N_{\rm bins}} \sum_{i,j,k,l=1}^{n_S} \left\langle \delta y_{a,i} \delta y_{c,k} \right\rangle \left\langle \delta y_{b,i} \delta y_{d,k} \right\rangle \left\langle \delta y_{a,j} \delta y_{c,l} \right\rangle \left\langle \delta y_{b,j} \delta y_{d,l} \right\rangle = \nonumber \\
& 4 \sum_{a,b,c,d=1}^{N_{\rm bins}} \sum_{i,j,k,l=1}^{n_S} \left( \delta^{ik} - \frac1{n_S} \right) \delta^{ac} \left( \delta^{ik} - \frac1{n_S} \right) \delta^{bd} \left( \delta^{jl} - \frac1{n_S} \right) \delta^{ac} \left( \delta^{jl} - \frac1{n_S} \right) \delta^{bd} = 4 N_{\rm bins}^2 (n_S-1)^2 \\
& 4 \sum_{a,b,c,d=1}^{N_{\rm bins}} \sum_{i,j,k,l=1}^{n_S} \left\langle \delta y_{a,i} \delta y_{c,k} \right\rangle \left\langle \delta y_{b,i} \delta y_{d,k} \right\rangle \left\langle \delta y_{a,j} \delta y_{d,l} \right\rangle \left\langle \delta y_{b,j} \delta y_{c,l} \right\rangle = \nonumber \\
& 4 \sum_{a,b,c,d=1}^{N_{\rm bins}} \sum_{i,j,k,l=1}^{n_S} \left( \delta^{ik} - \frac1{n_S} \right) \delta^{ac} \left( \delta^{ik} - \frac1{n_S} \right) \delta^{bd} \left( \delta^{jl} - \frac1{n_S} \right) \delta^{ad} \left( \delta^{jl} - \frac1{n_S} \right) \delta^{bc} = 4 N_{\rm bins} (n_S-1)^2 \\
& 4 \sum_{a,b,c,d=1}^{N_{\rm bins}} \sum_{i,j,k,l=1}^{n_S} \left\langle \delta y_{a,i} \delta y_{c,k} \right\rangle \left\langle \delta y_{b,i} \delta y_{c,l} \right\rangle \left\langle \delta y_{a,j} \delta y_{d,k} \right\rangle \left\langle \delta y_{b,j} \delta y_{d,l} \right\rangle = \nonumber \\
& 4 \sum_{a,b,c,d=1}^{N_{\rm bins}} \sum_{i,j,k,l=1}^{n_S} \left( \delta^{ik} - \frac1{n_S} \right) \delta^{ac} \left( \delta^{il} - \frac1{n_S} \right) \delta^{bc} \left( \delta^{jk} - \frac1{n_S} \right) \delta^{ad} \left( \delta^{jl} - \frac1{n_S} \right) \delta^{bd} = 4 N_{\rm bins} (n_S-1) \\
& 8 \sum_{a,b,c,d=1}^{N_{\rm bins}} \sum_{i,j,k,l=1}^{n_S} \left\langle \delta y_{a,i} \delta y_{c,k} \right\rangle \left\langle \delta y_{b,i} \delta y_{c,l} \right\rangle \left\langle \delta y_{a,j} \delta y_{d,l} \right\rangle \left\langle \delta y_{b,j} \delta y_{d,k} \right\rangle = \nonumber \\
& 8 \sum_{a,b,c,d=1}^{N_{\rm bins}} \sum_{i,j,k,l=1}^{n_S} \left( \delta^{ik} - \frac1{n_S} \right) \delta^{ac} \left( \delta^{il} - \frac1{n_S} \right) \delta^{bc} \left( \delta^{jl} - \frac1{n_S} \right) \delta^{ad} \left( \delta^{jk} - \frac1{n_S} \right) \delta^{bd} = 8 N_{\rm bins} (n_S-1) \\
& 4 \sum_{a,b,c,d=1}^{N_{\rm bins}} \sum_{i,j,k,l=1}^{n_S} \left\langle \delta y_{a,i} \delta y_{c,k} \right\rangle \left\langle \delta y_{b,i} \delta y_{d,l} \right\rangle \left\langle \delta y_{a,j} \delta y_{c,l} \right\rangle \left\langle \delta y_{b,j} \delta y_{d,k} \right\rangle = \nonumber \\
& 4 \sum_{a,b,c,d=1}^{N_{\rm bins}} \sum_{i,j,k,l=1}^{n_S} \left( \delta^{ik} - \frac1{n_S} \right) \delta^{ac} \left( \delta^{il} - \frac1{n_S} \right) \delta^{bd} \left( \delta^{jl} - \frac1{n_S} \right) \delta^{ac} \left( \delta^{jk} - \frac1{n_S} \right) \delta^{bd} = 4 N_{\rm bins}^2 (n_S-1)
\end{eqnarray}

Gathering all together gives
\begin{eqnarray} \label{eq:Var-tr-deltaY2}
\Var \Argf{\tr \Args{\Arg{\bf \delta Y}^2}} & \approx & \frac{N_{\rm bins}}{(n_S-1)^3} [8 N_{\rm bins}^2 + 16 N_{\rm bins} + 8 + 4 N_{\rm bins} (n_S-1) + 4 (n_S-1) + 4 + 8 + 4 N_{\rm bins}] = \nonumber \\
& & \frac{4 N_{\rm bins}}{(n_S-1)^3} [2 N_{\rm bins}^2 + 4 N_{\rm bins} + 4 + (N_{\rm bins} + 1) n_S] =
\frac{4 N_{\rm bins}}{(n_S-1)^3} [(N_{\rm bins} + 1) (n_S + 2 N_{\rm bins} + 2) + 2].
\end{eqnarray}

Then, according to Eq.~\eqref{eq:D_KL-Y-approx}, the variance of the KL divergence is approximately 16 times smaller:
\begin{eqnarray} \label{eq:D_KL-variance}
\Var [D_{\rm KL} \Arg{{\bf \Psi}_0, {\bf X}}] \approx \frac{N_{\rm bins} [(N_{\rm bins} + 1) (n_S + 2 N_{\rm bins} + 2) + 2]}{4(n_S-1)^3}.
\end{eqnarray}
This is an approximation because in Eq.~\eqref{eq:D_KL-Y-approx} the Taylor expansion was truncated at the quadratic order in $\bf \delta Y$.
Other derivation steps are exact.

\subsection{Inverse test}
\label{subsec:inv-test}

We considered how different $\chi^2$ would result from one matrix compared to the other.
If $\bm u$ is an unit vector in $N_{\rm bins}$-dimensional space ($\bm u^T \bm u = 1$), then $\bm w = {\bf C}_2^{1/2} \bm u$ gives an unit $\chi^2$ according to $C_2$: $\bm w^T {\bf C}_2^{-1} \bm w=1$.
Here ${\bf C}_2^{1/2}$ means the matrix square root of ${\bf C}_2$ -- a matrix with the same eigenvectors and eigenvalues equal to the square roots of corresponding eigenvalues of the original matrix.
Then we consider the $\chi^2$ with respect to ${\bf C}_1$ (${\bf \Psi}_1$): $\bm w^T {\bf \Psi}_1 \bm w$ and subtract the expected value of 1:
\begin{eqnarray}
\bm w^T {\bf \Psi}_1 \bm w - 1 = \bm u^T \Args{{\bf C}_2^{1/2} {\bf \Psi}_1 {\bf C}_2^{1/2} - {\bf \mathbb I}} \bm u.
\end{eqnarray}
Taking the RMS over all directions of $u$, one arrives at the RMS eigenvalue of this matrix, which can be expressed through the Frobenius norm:
\begin{eqnarray}
R_{\rm inv} \Arg{{\bf \Psi}_1, {\bf C}_2} = \frac1{\sqrt{N_{\rm bins}}} \Abs{\Abs{{\bf C}_2^{1/2} {\bf \Psi}_1 {\bf C}_2^{1/2} - {\bf \mathbb I}}}_F.
\end{eqnarray}
The Frobenius norm can be recast as a trace and simplified further using its cyclic property:
\begin{eqnarray}
\Abs{\Abs{{\bf C}_2^{1/2} {\bf \Psi}_1 {\bf C}_2^{1/2} - {\bf \mathbb I}}}_F &=& \tr \Args{\Arg{{\bf C}_2^{1/2} {\bf \Psi}_1 {\bf C}_2^{1/2} - {\bf \mathbb I}}^T \Arg{{\bf C}_2^{1/2} {\bf \Psi}_1 {\bf C}_2^{1/2} - {\bf \mathbb I}}} = \tr \Args{\Arg{{\bf C}_2^{1/2} {\bf \Psi}_1 {\bf C}_2^{1/2} - {\bf \mathbb I}}^2} \\ \nonumber
&=& \tr \Arg{{\bf C}_2^{1/2} {\bf \Psi}_1 {\bf C}_2 {\bf \Psi}_1 {\bf C}_2^{1/2} - 2 {\bf C}_2^{1/2} {\bf \Psi}_1 {\bf C}_2^{1/2} + {\bf \mathbb I}} = \tr \Arg{{\bf \Psi}_1 {\bf C}_2 {\bf \Psi}_1 {\bf C}_2 - 2 {\bf \Psi}_1 {\bf C}_2 + {\bf \mathbb I}} = \tr \tr \Args{\Arg{{\bf \Psi}_1 {\bf C}_2 - {\bf \mathbb I}}^2}.
\end{eqnarray}
This allows us to compute the quantity as
\begin{eqnarray}
R_{\rm inv} \Arg{{\bf \Psi}_1, {\bf C}_2} = \sqrt{\frac{\tr \Args{\Arg{{\bf \Psi}_1 {\bf C}_2 - {\bf \mathbb I}}^2}}{N_{\rm bins}}},
\end{eqnarray}
which is more computationally robust -- it is better to avoid factorizing (and inverting) covariance matrices (especially the sample one) when possible.

It is notable that this metric is related to the discriminant matrix
\begin{eqnarray} \label{eq:discriminant_matrix}
{\bf P} = \sqrt{{\bf \Psi}_R}^T {\bf C}_S \sqrt{{\bf \Psi}_R} - {\bf \mathbb{I}},
\end{eqnarray}
where $\sqrt{{\bf \Psi}_R}$ is the lower Cholesky decomposition, used in \cite{rascalC,rascalC-legendre-3}.
Through a similar procedure, one can find that its Frobenius norm is the same as above:
\begin{eqnarray}
\Abs{\Abs{\mathbf P}}_F^2 = \tr \Args{\Arg{{\bf \Psi}_1 {\bf C}_2 - {\bf \mathbb I}}^2}.
\end{eqnarray}
However, the interpretation of elements of the discriminant matrix is less clear.

Now let us consider the square of this metric (to remove the root):
\begin{eqnarray}
R_{\rm inv}^2 \Arg{{\bf \Psi}_0, {\bf X}} = \frac1{N_{\rm bins}} \tr [\Psi_0 X \Psi_0 X - 2 \Psi_0 X + \mathbb I].
\end{eqnarray}
Remembering Eq.~\eqref{eq:Y-relation}, we arrive to
\begin{eqnarray}
R_{\rm inv}^2 \Arg{{\bf \Psi}_0, {\bf X}} = \frac1{N_{\rm bins}} \tr \Args{{\bf Y}^2 - 2 {\bf Y} + {\bf \mathbb I}}.
\end{eqnarray}
Furthermore, expanding in $\bf \delta Y$ (according to Eq.~\eqref{eq:delta-Y}), we get
\begin{eqnarray} \label{eq:R_inv2-delta_Y}
R_{\rm inv}^2 \Arg{{\bf \Psi}_0, {\bf X}} = \frac1{N_{\rm bins}} \tr \Args{\Arg{\bf \delta Y}^2},
\end{eqnarray}
which is similar to Eq.~\eqref{eq:D_KL-Y-approx} up to a constant factor of $1/N_{\rm bins}$ and lack of approximations.
We can obtain the expectation value by plugging in Eq.~\eqref{eq:Y-covariance}:
\begin{eqnarray} \label{eq:R_inv2-expectation}
\Avg{R_{\rm inv}^2 \Arg{{\bf \Psi}_0, {\bf X}}} = \frac{N_{\rm bins}+1}{n_S-1}.
\end{eqnarray}
For variance we can use Eq.~\eqref{eq:Var-tr-deltaY2}:
\begin{eqnarray} \label{eq:R_inv2-variance}
\Var \Args{R_{\rm inv}^2 \Arg{{\bf \Psi}_0, {\bf X}}} = \frac{4 [(N_{\rm bins} + 1) (n_S + 2 N_{\rm bins} + 2) + 2]}{N_{\rm bins} (n_S-1)^3}.
\end{eqnarray}
Assuming $\Var \Args{R_{\rm inv}^2 \Arg{{\bf \Psi}_0, {\bf X}}} \ll \Avg{R_{\rm inv}^2 \Arg{{\bf \Psi}_0, {\bf X}}}^2$, we can take the square root to estimate the mean and variance of not-squared metric as
\begin{eqnarray} \label{eq:R_inv-expectation}
\Avg{R_{\rm inv} \Arg{{\bf \Psi}_0, {\bf X}}} & \approx & \sqrt{\Avg{R_{\rm inv}^2 \Arg{{\bf \Psi}_0, {\bf X}}}} = \sqrt{\frac{N_{\rm bins}+1}{n_S-1}}, \\ \label{eq:R_inv-variance}
\Var \Args{R_{\rm inv} \Arg{{\bf \Psi}_0, {\bf X}}} & \approx & \frac{\Var \Args{R_{\rm inv}^2 \Arg{{\bf \Psi}_0, {\bf X}}}}{4\Avg{R_{\rm inv}^2 \Arg{{\bf \Psi}_0, {\bf X}}}} = \frac{(N_{\rm bins} + 1) (n_S + 2 N_{\rm bins} + 2) + 2}{N_{\rm bins} (N_{\rm bins}+1) (n_S-1)^2}.
\end{eqnarray}
Also, it is useful to note that $R_{\rm inv}$ is related to the $\chi^2$ approximation of minus log-likelihood, considering the covariance of all covariance matrix elements.
This is rather easy to see with ${\bf \delta Y}$ expression (Eq.~\eqref{eq:R_inv2-delta_Y}).
Since ${\bf \delta Y}$ is real and symmetric, we obtain
\begin{eqnarray}
N_{\rm bins} \times R_{\rm inv}^2 \Arg{{\bf \Psi}_0, {\bf X}} = \Abs{\Abs{\bf \delta Y}}^2_F = \sum_{a,b=1}^{N_{\rm bins}} \Arg{\delta Y_{ab}}^2.
\end{eqnarray}
From Eq.~\eqref{eq:Y-covariance} we conclude that distinct elements of $\bf Y$ matrix have zero covariance (excluding the pairs symmetric with respect to the diagonal), its diagonal elements have a variance of $2/(n_S-1)$ and off-diagonal elements have a variance of $1/(n_S-1)$.
Then this covariance is trivial to invert and we have to sum squares of deviation of independent elements divided by their variance to get the $\chi^2$:
\begin{eqnarray}
\chi^2 = \Arg{n_S-1} \Args{\frac12 \sum_{a=1}^{N_{\rm bins}} \Arg{\delta Y_{aa}}^2 + \sum_{a=1}^{N_{\rm bins}} \sum_{b=a+1}^{N_{\rm bins}} \Arg{\delta Y_{ab}}^2} = \frac{n_S-1}2 \sum_{a,b=1}^{N_{\rm bins}} \Arg{\delta Y_{ab}}^2 = \frac{\Arg{n_S-1} N_{\rm bins}}2 \times R_{\rm inv}^2 \Arg{{\bf \Psi}_0, {\bf X}}.
\end{eqnarray}
Since elements of $\bf Y$ are independent linear combinations of elements of sample covariance matrix $\bf X$ (via Eq.~\eqref{eq:Y-relation}, since ${\bf \Psi}_0$ is not degenerate), the same holds for $\bf X$, but a direct computation without the ``rotation'' into $\bf Y$ would be significantly longer as the covariance of $X_{ab}$ elements (Eq.~\eqref{eq:X-covariance}) has a more generic and complex structure.

\subsection{Mean chi-squared}
\label{subsec:chi2}

We consider the sum of $\chi^2$ associated with the deviation of data vectors in individual samples from the estimate of average:
\begin{eqnarray}
\sum_{i=1}^{n_S} \sum_{a,b=1}^{N_{\rm bins}} (x_{a,i} - \bar x_a) \Psi_{0,ab} (x_{b,i} - \bar x_b) = (n_S-1) \sum_{a,b=1}^{N_{\rm bins}} \Psi_{0,ab} X_{ab} = (n_S-1) \tr (\Psi_0 X).
\end{eqnarray}
Remembering Eq.~\eqref{eq:measurement-normalization}, we can rewrite the LHS as
\begin{eqnarray} \label{eq:chi2_red_statistics}
\sum_{i=1}^{n_S} \sum_{a,b=1}^{N_{\rm bins}} (y_{a,i} - \bar y_a) \delta_{ab} (y_{b,i} - \bar y_b) = \sum_{a=1}^{N_{\rm bins}} \sum_{i=1}^{n_S} (y_{a,i} - \bar y_a)^2 \sim \sum_{a=1}^{N_{\rm bins}} \chi^2(n_S-1) \sim \chi^2[N_{\rm bins} \times (n_S-1)].
\end{eqnarray}
Therefore the corresponding reduced $\chi^2$ is
\begin{eqnarray}
\chi^2_{\rm red} = \frac1{N_{\rm bins}} \tr \Arg{{\bf \Psi}_0 {\bf X}}.
\end{eqnarray}

\subsection{Validation of means and standard deviations}
\label{subsec:stats-validation}

To check the theoretical results from the previous sections, we have performed a quick Monte-Carlo validation.
10,000 batches of 999 samples having 45 bins each have been generated.
For simplicity, we have taken the true covariance and precision to be unity matrices $C_0=\Psi_0=\mathbb I$.
This should not affect the results, save for numerical instabilities in matrix operations.
Full 45-bin sample covariance and precision matrices have been estimated in each batch.
Then, 25 bins were selected and projected into 5 quantities, for simplicity using 5 random orthonormal vectors as parameter derivatives.
Comparison between theoretical and sampled means and standard deviations are presented in Table~\ref{tab:stats-validation}.
Differences are most pronounced in $D_{\rm KL}$, but the disagreement is only in the second digit of standard deviation and fractions of standard deviation on the mean.
Therefore we report a close agreement, more than enough for the main part of the paper, where the scatter of \rascalc{} results is significantly larger than these standard deviations.
We note that the results for $D_{\rm KL}$ and $R_{\rm inv}$ (Eq.~\eqref{eq:D_KL-expectation}, \eqref{eq:D_KL-variance}, \eqref{eq:R_inv-expectation} and \eqref{eq:R_inv-variance}) are approximate and we expect meaningful deviations from them, especially as $N_{\rm bins}$ increases, while the derivations for $R_{\rm inv}^2$ and $\chi^2_{\rm red}$ (Eq.~\eqref{eq:R_inv2-expectation}, \eqref{eq:R_inv2-expectation} and \eqref{eq:chi2_red_statistics}) are exact.

\begin{table}
\begin{tabular}{ccccc}
\hline
 & & $D_{\rm KL}(\Psi_0, X)$ & $R_{\rm inv}(\Psi_0, X)$ & $\chi^2_{\rm red}(\Psi_0, X)$ \\
\hline
\multirow{2}{*}{Measurement space (45 bins)} & Theoretical & $0.519 \pm 0.024$ & $0.2147 \pm 0.0049$ & $1.0000 \pm 0.0067$ \\
 & Sampled & $0.526 \pm 0.023$ & $0.2146 \pm 0.0050$ & $1.0000 \pm 0.0067$ \\
\hline
\multirow{2}{*}{Parameter space (5 quantities projected from 25 bins)} & Theoretical & $0.0075 \pm 0.0028$ & $0.078 \pm 0.014$ & $1.000 \pm 0.020$ \\
 & Sampled & $0.0077 \pm 0.0028$ & $0.077 \pm 0.014$ & $1.000 \pm 0.020$ \\
\hline
\end{tabular}
\caption{Theoretical versus sampled mean $\pm$ std of covariance matrix comparison metrics with the true precision matrix. A close agreement can be seen.}
\label{tab:stats-validation}
\end{table}

We have repeated this test with a realistic covariance matrix (\rascalc{} Average NG for pre-reconstruction) and derivatives of the observables with respect to the parameters (accordingly, for the BAO model before reconstruction) to confirm whether the comparison measures are indeed not affected.
We have obtained the same numbers as in simpler test (Table~\ref{tab:stats-validation}), and close agreement of $\sigma\Args{\sigma(\alpha_{\rm BAO})}/\sigma(\alpha_{\rm BAO})$ with $\Args{2(n_S-1)}^{-1/2}$ according to Eq.~\eqref{eq:sigma-sigma-alpha}.


\section*{}

$^{1}$Center for Astrophysics | Harvard \& Smithsonian, 60 Garden St, Cambridge, MA 02138, USA \\
$^{2}$Lawrence Berkeley National Laboratory, 1 Cyclotron Road, Berkeley, CA 94720, USA \\
$^{3}$Department of Physics \& Astronomy, University College London, Gower Street, London, WC1E 6BT, UK \\
$^{4}$Institute for Computational Cosmology, Department of Physics, Durham University, South Road, Durham DH1 3LE, UK \\
$^{5}$Department of Physics and Astronomy, The University of Utah, 115 South 1400 East, Salt Lake City, UT 84112, USA \\
$^{6}$Instituto de F\'{\i}sica, Universidad Nacional Aut\'{o}noma de M\'{e}xico,  Cd. de M\'{e}xico  C.P. 04510,  M\'{e}xico \\
$^{7}$Center for Cosmology and AstroParticle Physics, The Ohio State University, 191 West Woodruff Avenue, Columbus, OH 43210, USA \\
$^{8}$Institut de F\'{i}sica d’Altes Energies (IFAE), The Barcelona Institute of Science and Technology, Campus UAB, E-08193 Bellaterra Barcelona, Spain \\
$^{9}$Departamento de F\'isica \& Observatorio Astron\'omico, Universidad de los Andes, Cra. 1 No. 18A-10, Edificio Ip, CP 111711, Bogot\'a, Colombia \\
$^{10}$Department of Astrophysical Sciences, Princeton University, Princeton NJ 08544, USA \\
$^{11}$Department of Physics, Southern Methodist University, 3215 Daniel Avenue, Dallas, TX 75275, USA \\
$^{12}$Instituci\'{o} Catalana de Recerca i Estudis Avan\c{c}ats, Passeig de Llu\'{\i}s Companys, 23, 08010 Barcelona, Spain \\
$^{13}$Department of Physics and Astronomy, Sejong University, Seoul, 143-747, Korea \\
$^{14}$Max Planck Institute for Astrophysics, Karl-Schwarzschild-Str. 1, D-85748 Garching, Germany \\
$^{15}$Institute of Cosmology \& Gravitation, University of Portsmouth, Dennis Sciama Building, Portsmouth, PO1 3FX, UK \\
$^{16}$National Astronomical Observatories, Chinese Academy of Sciences, A20 Datun Rd., Chaoyang District, Beijing, 100012, PR China \\
$^{17}$Space Sciences Laboratory, University of California, Berkeley, 7 Gauss Way, Berkeley, CA  94720, USA \\
$^{18}$CIEMAT, Avenida Complutense 40, E-28040 Madrid, Spain \\
$^{19}$Korea Astronomy and Space Science Institute, 776, Daedeokdae-ro, Yuseong-gu, Daejeon 34055, Republic of Korea \\
$^{20}$Department of Physics, University of Michigan, Ann Arbor, MI 48109, USA \\
$^{21}$Department of Physics \& Astronomy, Ohio University, Athens, OH 45701, USA \\
$^{22}$NSF's NOIRLab, 950 N. Cherry Ave., Tucson, AZ 85719, USA \\
$^{23}$Department of Astronomy, Tsinghua University, 30 Shuangqing Road, Haidian District, Beijing, 100190, China \\
$^{24}$Ecole Polytechnique F\'{e}d\'{e}rale de Lausanne, CH-1015 Lausanne, Switzerland


\bsp	
\label{lastpage}
\end{document}